\documentclass{aa}
\usepackage{graphicx}
\usepackage{txfonts}
\usepackage{mathrsfs}
\usepackage{color}
\usepackage{siunitx}

\begin{document}
\title{Hubble Space Telescope photometry of multiple stellar populations in the inner parts of NGC 2419}

\author{
    S{\o}ren S.\ Larsen
   \inst{1}
   \and
   Holger Baumgardt
   \inst{2}
   \and
   Nate Bastian
   \inst{3}
   \and
   Svea Hernandez
   \inst{1,4}
   \and
   Jean Brodie
   \inst{5}
}
\institute{
  Department of Astrophysics/IMAPP,
              Radboud University, PO Box 9010, 6500 GL Nijmegen, The Netherlands
  \and
  School of Mathematics and Physics, University of Queensland, St.\ Lucia, QLD 4072, Australia
  \and
  Astrophysics Research Institute, Liverpool John Moores University, 146 Brownlow Hill, Liverpool L3 5RF, United Kingdom    
  \and
  Space Telescope Science Institute, 3700 San Martin Drive, Baltimore, MD 21218, USA
  \and
  UCO/Lick Observatory, University of California, Santa Cruz, CA 95064, USA          
}

\offprints{S.\ S.\ Larsen, \email{S.Larsen@astro.ru.nl}}

\date{Received 23 October 2018 / Accepted 29 January 2019}

\abstract{
We present new deep imaging of the central regions of the remote globular cluster NGC~2419, obtained with the F343N and F336W filters of the Wide Field Camera~3 on board the \emph{Hubble Space Telescope}. The new data are combined with archival imaging to constrain nitrogen and helium abundance variations within the cluster.
We find a clearly bimodal distribution of the nitrogen-sensitive F336W-F343N colours of red giants, from which we estimate that about 55\% of the giants belong to a population with about normal (field-like) nitrogen abundances (P1), while the remaining 45\% belong to a nitrogen-rich population (P2). On average, the P2 stars are more He-rich than the P1 stars, with an estimated mean difference of $\Delta \mathrm{Y} \simeq0.05$, but the P2 stars exhibit a significant spread in He content and some may reach $\Delta \mathrm{Y}\simeq0.13$. A smaller He spread may also be present for the P1 stars.
Additionally, stars with spectroscopically determined low Mg abundances ($\mathrm{[Mg/Fe]}<0$)  are generally associated with P2.
We find the P2 stars to be slightly more centrally concentrated in NGC~2419 with a projected half-number radius of about 10\% less than for the P1 stars, but the difference is not highly significant ($p\simeq0.05$).
Using published radial velocities, we find evidence of rotation for the P1 stars, whereas the results are inconclusive for the P2 stars, which are consistent with no rotation as well as the same average rotation found for the P1 stars.
Because of the long relaxation time scale of NGC~2419, the radial trends and kinematic properties of the populations are expected to be relatively unaffected by dynamical evolution. 
Hence, they provide constraints on formation scenarios for multiple populations, which must account not only for the presence of He spreads within sub-populations identified via CNO variations, but also for the relatively modest differences in the spatial distributions and kinematics of the populations.
}


\titlerunning{NGC 2419}
\maketitle

\section{Introduction}

In the words of \citet{Shapley1922}, NGC~2419 is `a globular cluster of uncommon interest'. At a distance of 83~kpc \citep{Ripepi2007}, it has long been recognised as one of the most remote globular clusters (GCs) in the Milky Way \citep{Baade1935,Racine1975}. Despite the difficulties associated with the great distance, NGC~2419 has a number of unique characteristics that motivate the effort required to undertake detailed studies.
It is among the most luminous and massive GCs in the Milky Way and it is also very extended; these properties together imply a very long relaxation time. \citet{Baumgardt2018} found a total mass of $(9.8\pm1.4)\times10^5 M_\odot$ and a
half-mass radius of $r_h = 24.2$~pc, corresponding to a half-mass relaxation time of $t_\mathrm{rh} = 55 \, \mathrm{Gyr}$. This is by far the longest among the GCs in the Milky Way. In many GCs, the present-day structural parameters have likely been significantly modified by dynamical effects such as mass segregation and orbital mixing \citep{Decressin2008,Vesperini2013,Dalessandro2014,Dalessandro2018}. Because of the long relaxation time, such effects are expected to be minimal in NGC~2419 and it therefore offers one of the best opportunities to constrain the initial structural properties of a globular cluster.

The vast majority of GCs that have been studied in detail to date exhibit variations in the abundances of many of the light elements, with anti-correlated abundances of Na/O, C/N, and, less commonly, Al/Mg
\citep{Carretta2009,Cohen2002,Shetrone1996,Sneden2004}.  
NGC~2419 is no exception, and some of its abundance variations are in fact more extreme than in most other GCs: about half of the stars in NGC~2419 have extremely depleted Mg abundances, reaching $\mathrm{[Mg/Fe]}$ values as low as $-1$, and these stars are also enriched in potassium \citep{Cohen2012,Mucciarelli2012a}. However, the overall metallicity spread appears to be very small (less than $\sim$0.1 dex), with a mean metallicity of $\mathrm{[Fe/H]} = -2.09$ \citep{Mucciarelli2012a,Frank2015}.

Photometric studies reveal an extremely extended blue tail of the horizontal branch in NGC~2419 \citep{Ripepi2007,Dalessandro2008,Sandquist2008}, which suggests the presence of a population of strongly He-enriched stars \citep[$\mathrm{Y}\approx0.36$;][]{DiCriscienzo2011,DiCriscienzo2015}. 
\citet{DiCriscienzo2011} also noted a significant spread in the F435W-F814W ($\sim B\!-\!I$) colours of red giant branch (RGB) stars, and estimated that the colour spread was consistent with about 30\% of the stars being He-enriched. A spread in the colours of RGB stars was also found by \citet[][hereafter B2013]{Beccari2013} from ground-based $uVI$ photometry. According to B2013, stars with blue $u\!-\!V$ and $u\!-\!I$ colours tended to be more centrally concentrated within NGC~2419 than those with redder colours, and the authors argued that the blue colours were indicative of enhanced He. A more centrally concentrated distribution of stars with anomalous (i.e., non field-like) abundances of He and other elements would indeed be expected in some self-enrichment scenarios for the origin of multiple populations (MPs) in GCs \citep[e.g.][]{DErcole2008,Decressin2008,Bastian2013a}. However, B2013 also found that  stars with anomalous (low) Mg abundances tended to have redder $u\!-\!V$ colours than those with normal Mg abundances, in apparent conflict with the expectation that elevated He content would be coupled with depleted Mg abundances.
A complication when interpreting the $uVI$ colours of red giants is that the SDSS $u$-band is also sensitive to N abundance variations. An increased amount of N will suppress the flux in the $u$-band and lead to redder $u\!-\!V$ colours compared to a N-normal population. The net effect on the colours of an `enriched' population thus depends on the balance between the opposing effects of He- and N variations. This balance may change as a function of luminosity on the RGB. In this context, it is worth noting that the radial distributions were determined by B2013 for stars on the lower RGB, whereas the stars with spectroscopic Mg abundance measurements are found near the tip of the RGB. It is thus important to establish how the Mg abundance anomalies correlate with variations in other elemental abundances by independently constraining the He and N abundances of the different populations.

The most common spectroscopic tracers of multiple populations in GCs are the Na/O \citep{Carretta2009} and C/N \citep{Cohen2002} anti-correlations. Variations in CNO abundances are detectable photometrically through their effects on the OH, CN, NH, and CH molecular absorption bands in the blue part of the spectra of cool stars \citep[e.g.][]{Sbordone2011,Carretta2011}, as demonstrated in spectacular fashion by the Hubble Space Telescope (HST) UV Legacy Survey of globular clusters \citep{Piotto2015,Milone2017}. However, the distance of NGC~2419 makes UV observations of RGB stars relatively time consuming. 
From ground-based Str{\"o}mgren $uvby$ photometry, \citet[][hereafter F2015]{Frank2015} found a significant spread in N abundance for RGB stars in the outer parts of NGC~2419, with an approximately equal mix of two populations with distinct N abundances being favoured over a single Gaussian distribution of N abundances. This is reminiscent of the bimodality observed in the $\mathrm{[Mg/Fe]}$ and $\mathrm{[K/Fe]}$ abundance ratios \citep{Mucciarelli2012a}. The Str{\"o}mgren colours were found by F2015 to be consistent with the Mg-normal stars being N-normal, and Mg-poor stars being N-rich.

The inner regions of NGC~2419 are too crowded for accurate ground-based photometry of all but the brightest RGB stars, and the studies of B2013 and F2015 were both restricted to stars outside the central $\sim$50\arcsec\  (about one half-light radius) of the cluster. F2015 pointed out that the difference between the roughly equal fractions of N-normal and N-rich stars found by them in the outer regions of the cluster and the somewhat smaller fraction of He-rich stars in the centre \citep{DiCriscienzo2011} might suggest an inverse population gradient, in the sense that the enriched stars are somewhat less concentrated. While this might be at odds with theoretical expectations, it would not be completely unprecedented. 
In \citet{Larsen2015}, it was found that stars with enhanced N abundances were less concentrated than those with normal N abundances within the central regions of the GC M15, whereas an apparent reversal of this trend occurred at larger radii. However, this result has recently been challenged by \citet{Nardiello2018}, who found no significant difference in the radial distributions of the different populations in M15.

The correspondence between the stellar populations identified in the central regions of NGC~2419 (via the extended HB and the spread in optical colours on the RGB) and the constraints on N abundance variations in the outer parts remains unclear. What is still missing is a more robust way of establishing the contributions of N-abundance vs.\ He abundance variations to the colour variations in the central regions of the cluster.
HST observations in the F275W filter used by \citet{Piotto2015} would require impractically long integration times, but a viable alternative is offered by the narrow-band F343N filter, which is sensitive to the NH absorption band near 3400~\AA\ \citep{Larsen2014a}. Here we present new HST observations of NGC~2419 in the F343N and F336W filters, which we combine with existing archival data in several optical filters to constrain the properties of multiple populations in the inner regions of the cluster. 

Throughout this paper we assume a distance modulus of $(m-M)_0 = 19.60$~mag \citep{Ripepi2007} and a foreground extinction in the HST filters of 
$A_\mathrm{F336W} = 0.271$ mag, 
$A_\mathrm{F438W} = 0.220$ mag, 
$A_\mathrm{F555W} = 0.174$ mag, 
$A_\mathrm{F606W} = 0.151$ mag, 
$A_\mathrm{F814W} = 0.093$ mag, and
$A_\mathrm{F850LP} = 0.073$ mag \citep[][via the NASA/IPAC Extragalactic Database, NED]{Schlafly2011}.
Because of the low concentration of NGC~2419, the exact location of the centre is uncertain by several arcsec.
The 2010 edition of the \citet{Harris1996} catalogue gives the J2000.0 centre coordinates as (RA, Decl) = (07h38m08.47s,  $+38^\circ52^\prime56\farcs8$) whereas NED lists the coordinates as (07h38m07.9s, $+38^\circ52^\prime48\arcsec$). \citet{Dalessandro2008} used HST photometry to compute the location of the barycentre as (07h38m08.47s,  $+38^\circ52^\prime55\arcsec$). From looking at our new HST images, the Harris coordinates seem a little too far north, and those given by the NED too far south. We thus adopted the coordinates from \citet{Dalessandro2008}.

\section{Observations and data reduction}

\subsection{The WFC3 filters and multiple populations}

   \begin{figure*}
   \centering
   \includegraphics[width=16cm]{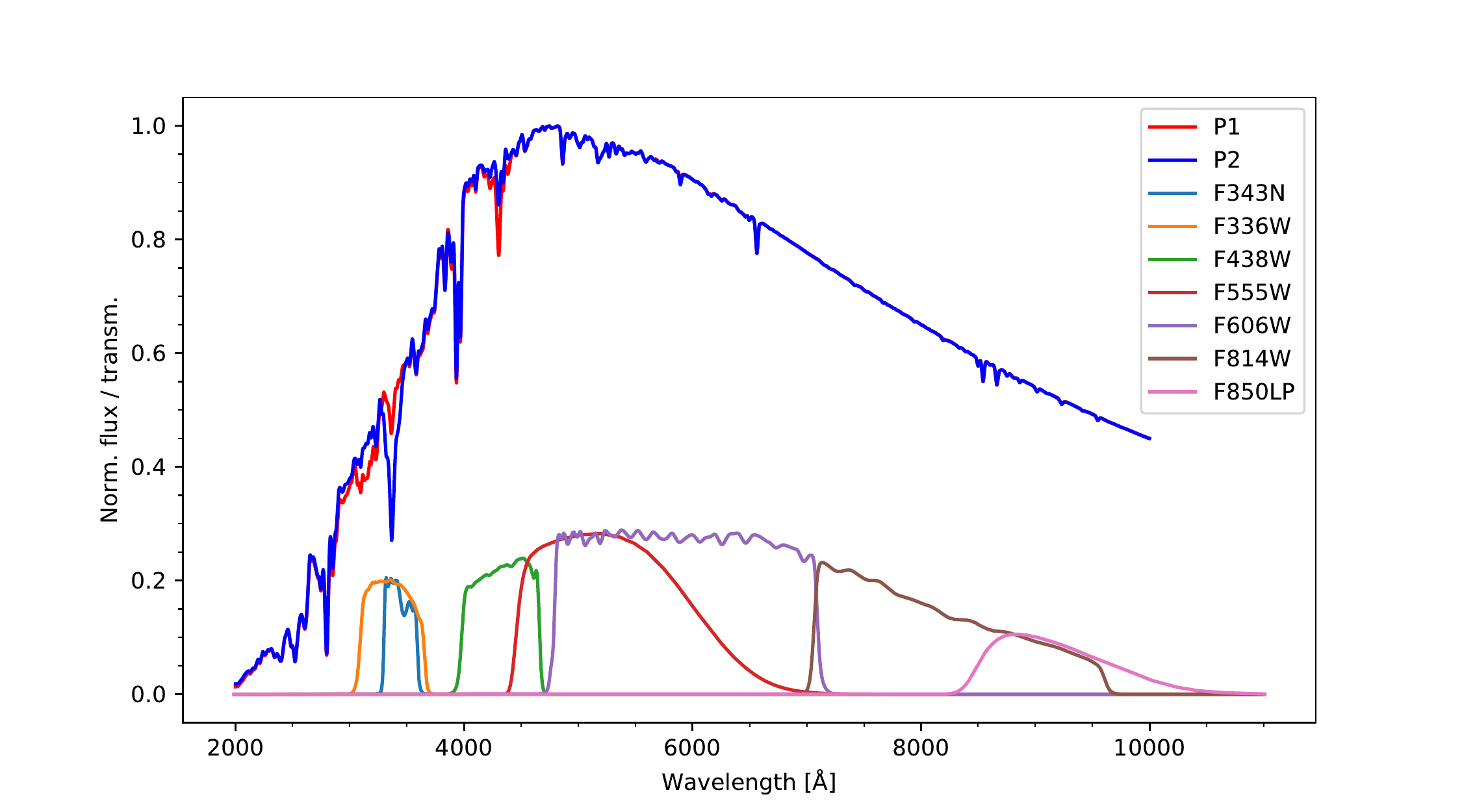}
      \caption{Model spectra of RGB stars with P1 and P2-like composition. Both model spectra are computed for $T_\mathrm{eff}=5254$ K, $\log g = 2.76$, and $\mathrm{[Fe/H]}=-2.0$.
      Also shown are the transmission curves for the filters used in this paper.
         \label{fig:p1p2spec}
         }
  \end{figure*}

The sensitivity of various photometric systems to multiple populations in GCs has been discussed in detail by previous studies \citep[e.g.][]{Carretta2011,Sbordone2011,Piotto2015}. The photometric signatures can be grouped into two broad categories: 1) atmospheric effects, and 2) effects on stellar structure. In the first case, the observed spectral energy distribution (SED) is modified by variations in the strengths of strong molecular absorption features (CN, CH, NH, OH), which are linked to variations in the light-element abundances  (C, N, O). As long as the C+N+O sum is constant, these abundance variations are not expected to have any significant effect on the structure of the star itself. In the second case, the abundance variations (typically an increased He content) do affect the structure of the star, and a He-enriched isochrone will generally be shifted to higher effective temperatures (bluer colours). 

Figure~\ref{fig:p1p2spec} shows model spectra for two stars with properties similar to those found near the base of the RGB in NGC~2419 ($T_\mathrm{eff}=5254$ K, $\log g = 2.76$, and $\mathrm{[Fe/H]}=-2.0$). The model atmospheres and corresponding spectra were computed with the \texttt{ATLAS12} and \texttt{SYNTHE} codes \citep{Sbordone2004,Kurucz2005} for normal ($\alpha$-enhanced) halo-like composition (P1) and the CNONa2 mixture of \citet{Sbordone2011}, which is typical of enriched stars (P2) in GCs  ($\Delta$(C, N, O, Na) = $-0.6, +1.44, -0.8, +0.8$ dex). Also included are the transmission curves for the WFC3 filters used in this paper. We see that the F343N filter samples the NH feature near 3400~\AA, which is much stronger in the N-rich P2 star. The broader F336W filter is also sensitive to variations in the OH bands bluewards of the NH feature, which are weaker in the P2 spectrum (due to the depleted O abundance). To the extent that stars in NGC~2419 follow the usual tendency for N and O to be anti-correlated, this enhances the ability of the F336W-F343N colour to distinguish between the different populations. We note also that the F438W filter includes the CH band near 4300~\AA\ (the Fraunhofer G feature), which is weaker in the P2 spectrum because of the C depletion. The F555W, F606W, F814W, and F850LP filters include no strong molecular bands and are largely insensitive to the atmospheric effects of multiple populations, but they are, of course, sensitive to variations in effective temperature that may be caused by variations in He content.

\subsection{Observations}

For the optical photometry we used archival data from observing programme GO-11903 (P.I.: J.\ Kalirai), which imaged the central parts of NGC~2419 with the Wide Field Camera 3 (WFC3) on board HST. These observations consist of pairs of un-dithered exposures in many filters. In this paper we use observations in F438W (exposure times of $2\times725$ s), F555W ($2\times580$ s), F606W (2$\times400$ s), F814W ($2\times650$ s), and F850LP ($2\times675$ s). The programme also includes exposures in several ultraviolet filters \citep[which were used by][]{DiCriscienzo2015}, but these are generally short and do not allow us to reach the required photometric accuracy for stars on the RGB. 

Additional observations in F336W and F343N were obtained in cycle 25 under programme GO-15078 (P.I.: S.\ Larsen). 
This filter combination was chosen specifically to measure variations in the strength of the NH band near 3400~\AA\ (Fig.~\ref{fig:p1p2spec}). The magnitude difference between the two filters is sensitive to the strength of the NH feature in a manner similar to, for example, the Str{\"o}mgren $\beta$ index for the H$\beta$ line \citep{Stromgren1966}.

The GO-15078 observations consist of three visits, of which two visits were allocated to F343N imaging and one visit to F336W. Each visit had a duration of three orbits and within each visit, the observations were dithered according to the C6 $3\times2$ dither pattern described in \citet{Dahlen2010}. Hence, NGC~2419 was observed in F343N for six orbits which yielded 12 exposures with a total exposure time of 17152 s. In F336W, the six exposures obtained during three orbits had a total exposure time of 8576 s. 
The roll angles, centre coordinates, and dither patterns were identical for all three visits. Given that the two filters have very similar central wavelengths, we expect that most systematic effects (reddening, sensitivity, etc.) will cancel out when calculating the difference between the F336W and F343N magnitudes. 

\subsection{Photometry}

For the analysis we used the `\texttt{*\_flc}' frames, which are corrected for charge-transfer inefficiencies by the instrument pipeline \citep{Ryan2016}.
Photometry was carried out with \texttt{ALLFRAME} \citep{Stetson1994}, following the procedure described in \citet{Larsen2014a}. 
After cleaning the individual pipeline-reduced frames of cosmic rays and multiplying them by the appropriate pixel area maps, \texttt{ALLFRAME} was set up to carry out photometry on each individual frame. Point-spread functions (PSFs) were determined from about 50 isolated, bright stars distributed evenly across each detector.
We followed the standard approach of detecting stars with the \texttt{find} task in \texttt{daophot} and carrying out a first round of aperture- and PSF-fitting photometry, then detecting additional stars on the star-subtracted images generated by \texttt{ALLFRAME} in the first pass, and using the merged star catalogues as input for a second pass of PSF-fitting photometry \citep{Stetson1987}. The PSFs used in the second pass were redetermined from images in which all stars except the PSF stars had been subtracted. 

The GO-15078 and GO-11903 data were reduced separately and the photometry catalogues were then merged. The photometry was calibrated to STMAG magnitudes using aperture photometry of the PSF stars and the 2017 photometric zero-points for an $r=10$ pixels aperture published on the WFC3 webpage\footnote{\url{http://www.stsci.edu/hst/wfc3/analysis/uvis_zpts/}}.
These are: $z_\mathrm{F343N} = 22.770$ mag, $z_\mathrm{F336W} = 23.517$ mag, $z_\mathrm{F438W} = 24.236$ mag, $z_\mathrm{F555W} = 25.651$ mag, $z_\mathrm{F606W} = 26.154$ mag, $z_\mathrm{F814W} = 25.861$ mag, and $z_\mathrm{F850LP} = 24.885$ mag.

Conveniently, the roll angles of the two datasets differ by less than 10 degrees (as indicated by the header keyword PA$\_$V3, which has a value of 276 deg for the GO-15078 observations and 269 deg for the GO-11903 observations) and the difference between the centre coordinates is only about $10\arcsec$. Hence, the overlap between the datasets is excellent. We used the \texttt{geomap} and \texttt{geoxytran} tasks in the \texttt{images.immatch} package in IRAF to define coordinate transformations between the GO-15078 and GO-11903 datasets.
Because of the good overlap, most of the stars imaged on a given detector in GO-15078 were mapped onto the same detector in GO-11903, although a small fraction of CCD\#1 in GO-15078 was mapped onto CCD\#2 in GO-11903 and vice versa. We used about 100--150 stars on each CCD to define the transformations, which were fitted with 3rd order polynomials in the $x$ and $y$ coordinates. This yielded an r.m.s.\ scatter of about 0.05 pixels in the solutions.
The pixel coordinates (measured in the GO-15078 frames) were further transformed to sky coordinates using the \texttt{wcs} package in \texttt{Astropy} \citep{AstropyCollaboration2018}. Small offsets to the HST coordinates (about 0\farcs216 in right ascension and 0\farcs061 in declination) were applied in order to match the Gaia astrometry, based on about 400 stars in common between our HST data and the Gaia DR2 catalogue \citep{GaiaCollaboration2016,GaiaCollaboration2018}. The dispersions around the mean offsets were about $0\farcs026$ and $0\farcs015$ in right ascension and declination, respectively, from which we estimate the remaining systematic uncertainty on the astrometric calibration to be about $10^{-3}$ arcsec.
The first few entries of the photometric catalogue are listed in Table~\ref{tab:photometry}, and the full catalogue is available on-line.

   \begin{figure}
   \centering
    \includegraphics[width=\columnwidth]{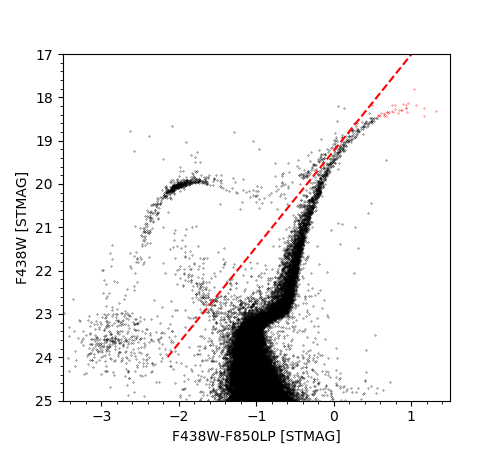}
      \caption{($m_\mathrm{F438W}-m_\mathrm{F850LP}, m_\mathrm{F438W}$) colour-magnitude diagram. Red symbols indicate stars that are saturated in the F850LP filter.
         \label{fig:cmd_bl_b}
         }
   \end{figure}
   
   \begin{figure}
   \centering
   \includegraphics[width=\columnwidth]{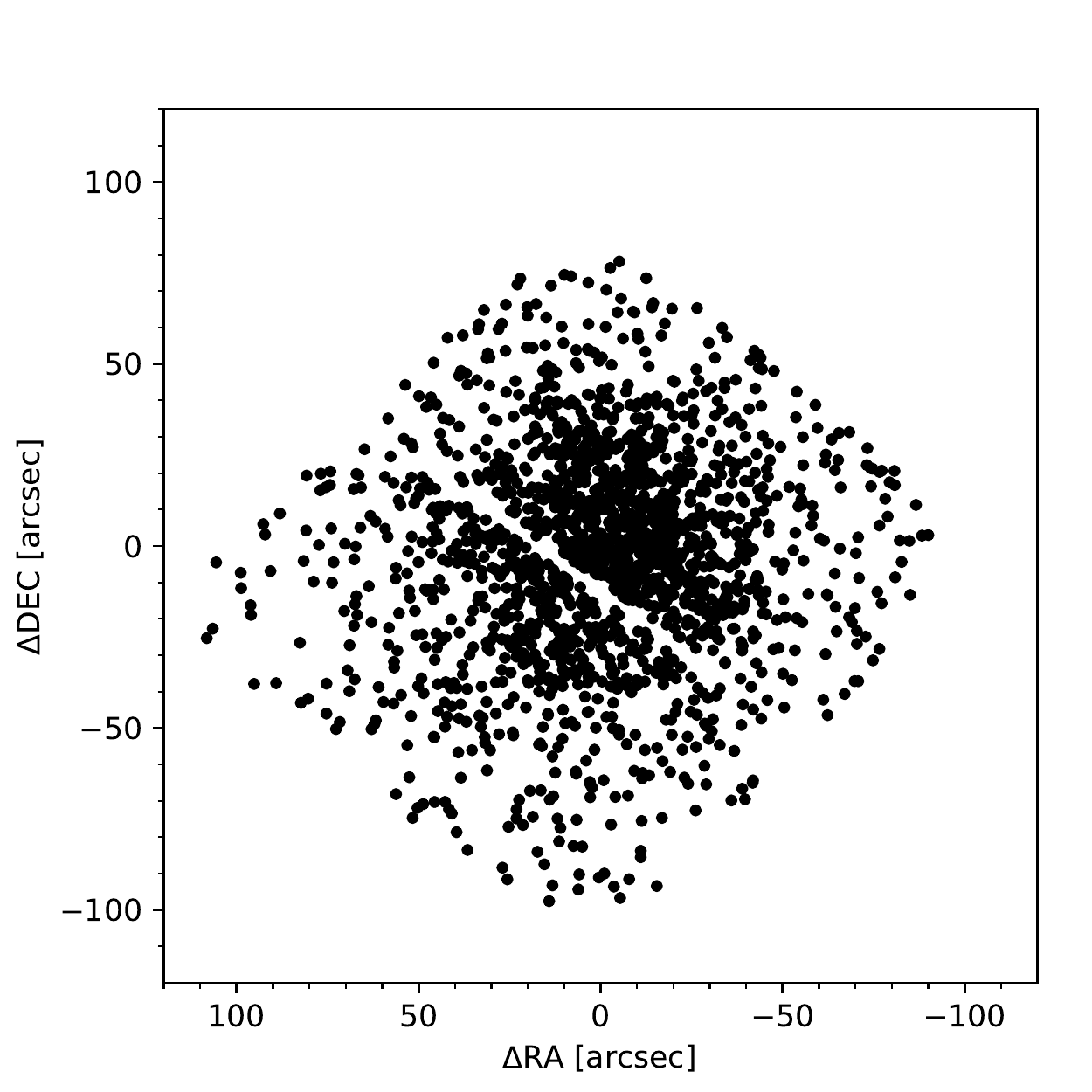}
      \caption{Spatial distribution of RGB stars relative to the centre of NGC~2419.
         \label{fig:map}
         }
   \end{figure}

Figure~\ref{fig:cmd_bl_b} shows the ($m_\mathrm{F438W}$ vs.\ $m_\mathrm{F438W-F850LP}$) colour-magnitude diagram (CMD). Overall, the CMD is very similar to that shown in Fig.~1 of \citet{DiCriscienzo2015}, apart from an offset due to our use of STMAG (instead of VEGAMAG) zero-points. The CMD clearly shows the main features identified in previous studies, including the horizontal branch (HB) and its extended blue tail, a population of blue stragglers (BS), as well as the relatively narrow RGB. The photometry reaches a couple of magnitudes below the main sequence turn-off (MSTO), although we will concentrate exclusively on the RGB in this paper. 

The red dashed line indicates the colour cut that we will use to separate RGB stars from potential HB and BS interlopers, which is given by: $m_\mathrm{F438W}-m_\mathrm{F850LP} > -0.12  - 0.45 \times (m_\mathrm{F438W} - 19.5)$.
We use the F438W-F850LP colour combination for this purpose, partly because it offers a long colour baseline, which helps to separate the RGB from AGB stars in the range $19 \la m_\mathrm{F438W} \la 20$, and partly because the F850LP filter is less affected by saturation than the F555W and F814W filters. In F555W and F814W, saturation sets in at $m_\mathrm{F555W} \sim 18.9$ and $m_\mathrm{F814W} \sim 19.0$, respectively, whereas the saturation limit is about one magnitude brighter, at $m_\mathrm{F850LP} \sim 17.9$  in F850LP. 
Stars that are saturated in F850LP are indicated with red symbols in Fig.~\ref{fig:cmd_bl_b}. 
In the F336W, F343N, and F438W observations, even the brightest RGB stars remain unsaturated. 
While it is possible to recover the flux accurately even for stars that are several magnitude brighter than the saturation limit \citep{Gilliland2010}, we have not attempted to do so here.

Figure~\ref{fig:map} shows the spatial distribution of RGB stars brighter than $M_\mathrm{F438W}=+2$ that are included in both the GO-15078 and GO-11903 datasets, relative to the adopted centre of NGC~2419. The coverage is spatially complete within a radius of $\approx70\arcsec$ (apart from the gap between the WFC3 detectors), and the outermost star is about 110\arcsec\ from the centre. In terms of the projected half-light radius \citep[$r_{h,lp}=45\arcsec$;][]{Baumgardt2018}, spatial coverage is thus complete to about $1.5 \, r_{h,lp}$.

\subsection{Artificial star tests}
\label{sec:artstar}

   \begin{figure}
   \centering
   \includegraphics[width=\columnwidth]{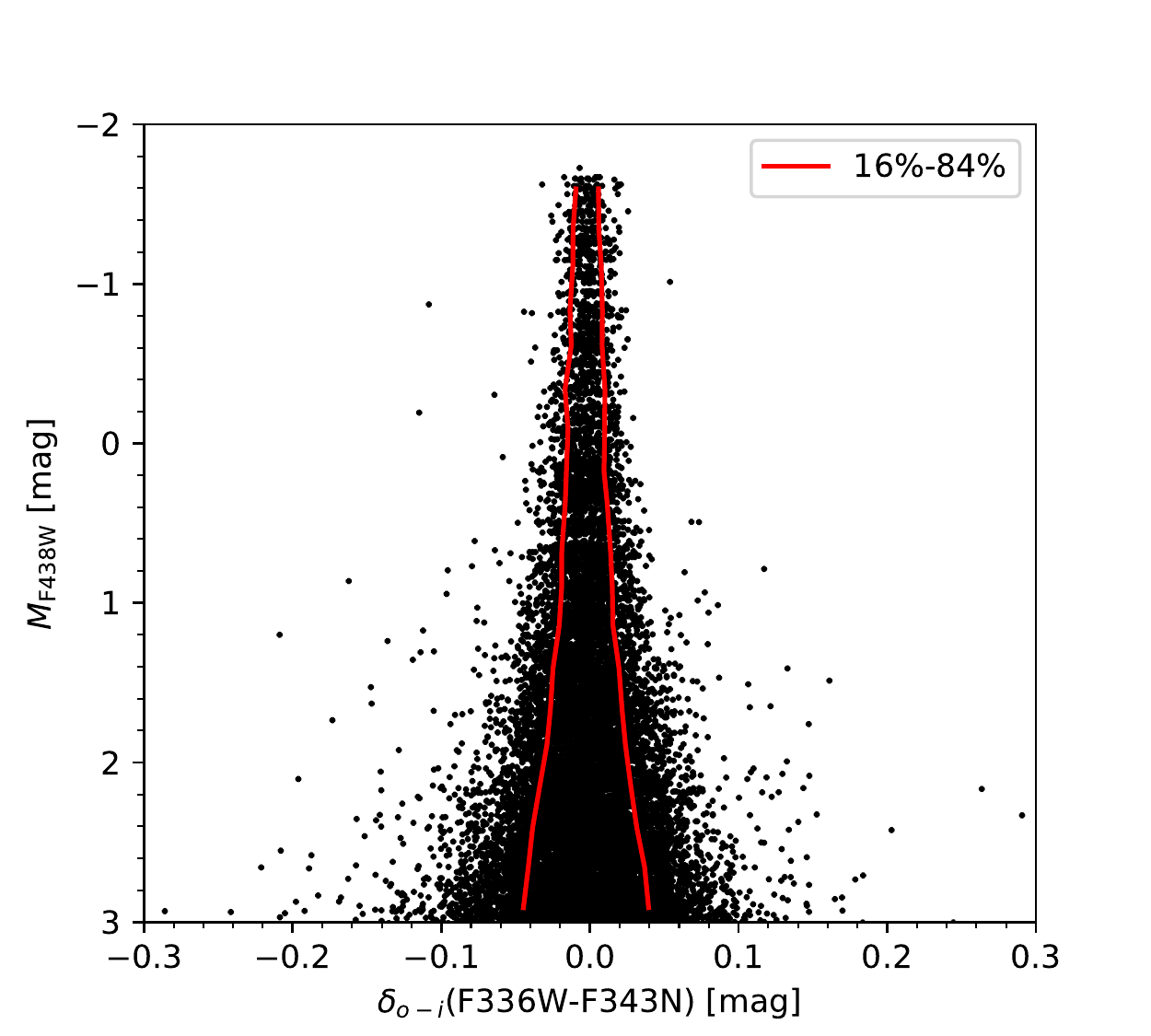}
      \caption{Input F438W magnitudes versus the difference between input and recovered F336W-F343N colour, $\delta_{o-i}$(F336W-F343N), for the artificial stars. The red lines show the 16\% and 84\% percentiles.  
         \label{fig:synt_dunu_b}
         }
   \end{figure}

   \begin{figure}
   \centering
   \includegraphics[width=\columnwidth]{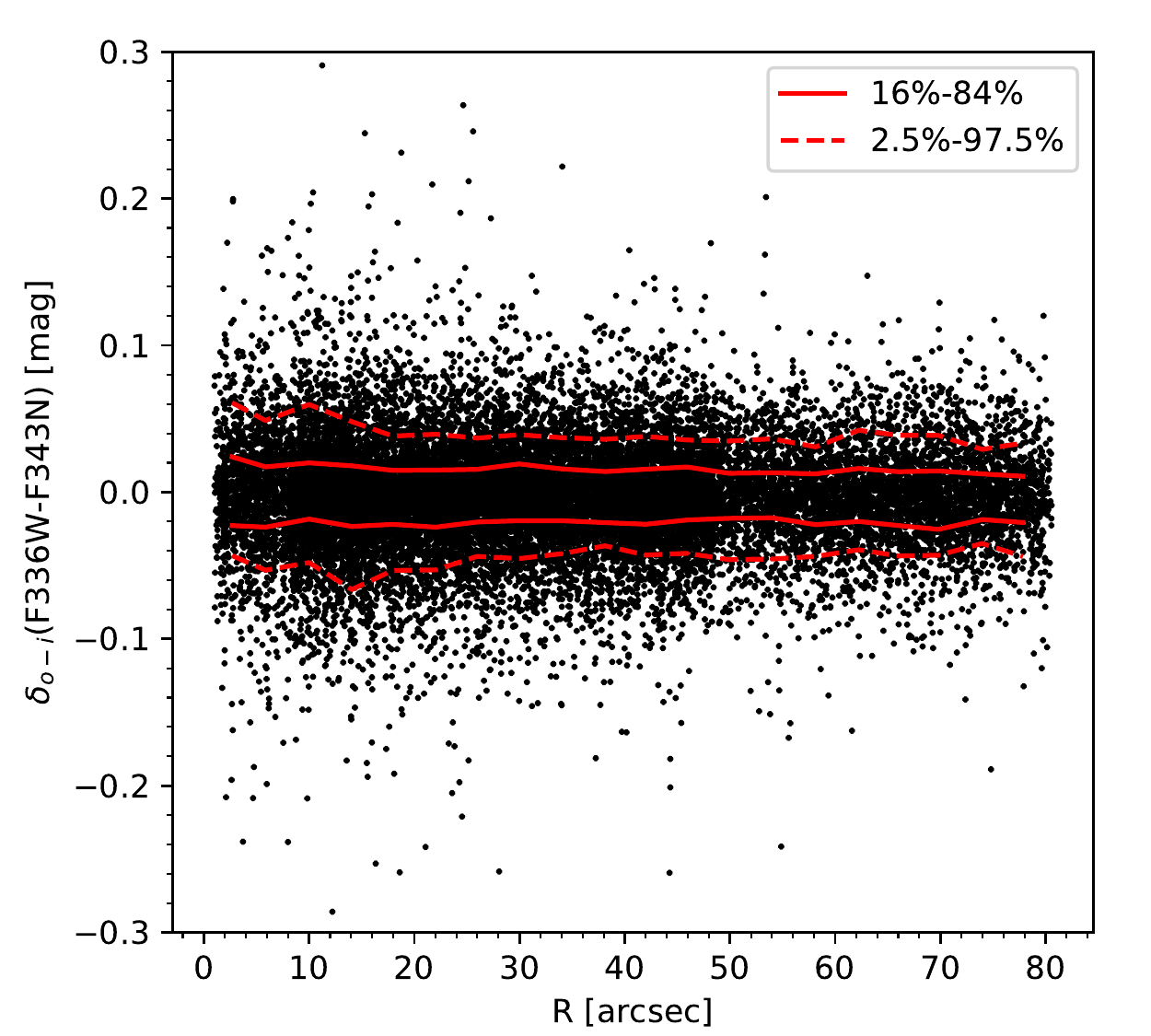}
      \caption{Here $\delta_{o-i}$(F336W-F343N) is plotted as a function of radial distance $R$ from the centre of NGC~2419 for artificial stars brighter than $M_\mathrm{F438W}=+2$. The red solid (dashed) lines show the 16\% and 84\% (2.5\% and 97.5\%) percentiles.  
         \label{fig:synt_r_dunu}
         }
   \end{figure}

The photometric accuracy and completeness were quantified by means of artificial star tests. Ten rounds of such tests were carried out, adding about 2050 stars to the HST images in each round with the \texttt{mksynth} task in the \texttt{BAOLAB} package \citep{Larsen1999}. 
The \texttt{mksynth} task models artificial stars by treating the PSF as a probability density function, from which events are picked at random and added to the image one by one until the desired number of counts has been reached. In this way, the images of simulated stars are subject to the same stochastic effects as those of real stars. 

To ensure that the central regions of the cluster were well sampled, the artificial stars were grouped into a series of concentric annuli, centred on NGC~2419. These annuli had radii of $0 < r_1 < 100$ pixels (60 stars per round), $100 < r_2 < 200$ pixels (90 stars), $200 < r_3 < 400$ pixels (325 stars), $400 < r_4 < 600$ pixels (300 stars), $600 < r_5 < 800$ pixels (280 stars), $800 < r_6 < 1200$ pixels (540 stars), and $1200 < r_7 < 2000 $ pixels (480 stars).  To avoid self-crowding among the artificial stars, we enforced a minimum separation of 20 pixels between any pair of artificial stars. A set of dedicated artificial PSF stars were also added. 

The F814W magnitudes of the artificial stars were picked at random from the actual magnitude distribution of RGB stars in NGC~2419, with the faintest artificial stars having an absolute magnitude of $M_\mathrm{F814W} = +4$ ($M_\mathrm{F438W} \approx +3$). To determine the magnitudes in the other filters, we used \texttt{ATLAS12} and \texttt{SYNTHE} to compute synthetic spectra for 20 sampling points along the RGB of an $\alpha$-enhanced isochrone with $\mathrm{[Fe/H]}=-2$ and an age of 13 Gyr \citep{Dotter2007}. Magnitudes in the STMAG system were then determined by integrating the synthetic spectra over the filter transmission curves, and for a given $M_\mathrm{F814W}$ magnitude the magnitudes in other filters were then obtained by interpolation in the synthetic relations.

The \texttt{ALLFRAME} procedure was then repeated, the only modification with respect to the original photometry being that the PSFs were determined from the artificial PSF stars. This was done to ensure that inaccuracies in the PSF determination were propagated properly through the procedure, rather than fitting the artificial stars directly with the same PSFs that were used to generate them in the first place. 
The resulting photometry catalogues were matched against the input lists of artificial stars. An artificial star in the input catalogue was defined as recovered if a counterpart was found within a distance of 1 pixel in the \texttt{ALLFRAME} catalogue.
The artificial star experiments were carried out separately for the GO-15078 and GO-11903 data, but using the same input catalogue of artificial stars. The photometry catalogues for the artificial stars could then subsequently be merged in the same way as was done for the science data. 

Essentially all of the artificial stars added to the images were recovered by the \texttt{ALLFRAME} photometry procedure in both the GO-15078 and GO-11903 datasets. Even at the faintest magnitudes, about one magnitude below the limits adopted in our analysis, the recovery fraction remained at $>99\%$ at all radii. Hence, we can assume that the \texttt{ALLFRAME} photometry is unaffected by completeness effects over the magnitude range of interest here. 

In Figure~\ref{fig:synt_dunu_b} we plot the input F438W magnitude vs. the difference between the input and recovered (output) F336W-F343N colour, $\delta_{o-i}$(F336W-F343N). The red curves indicate the 16\% and 84\% percentiles of $\delta_{o-i}$(F336W-F343N), computed in 0.25 mag bins. As expected, the scatter increases towards the faint end of the magnitude distribution. Taking half of the separation between the 16\% and 84\% percentiles as an estimate of the standard error, $\sigma_\mathrm{F336W-F343N}$, we find that this increases from $\sigma_\mathrm{F336W-F343N} = 0.008$ mag at the bright end to $\sigma_\mathrm{F336W-F343N} = 0.028$ mag at $M_\mathrm{F438W} = +2$, which will be the typical faint limit adopted in most of the subsequent analysis.

While crowding does not appear to affect our ability to detect RGB stars even at the centre of NGC~2419, it may still have an effect on the photometric errors.  Figure~\ref{fig:synt_r_dunu} shows $\delta_{o-i}$(F336W-F343N) as a function of distance from the centre of NGC~2419 for stars brighter than $M_\mathrm{F438W} = +2$. Plots for other colours look very similar. 
We see that there is indeed a mild tendency for the scatter to increase towards the centre. Estimating the standard error in the same way as above, we find $\sigma_\mathrm{F336W-F343N} = 0.023$~mag in the innermost bin ($\langle R \rangle = 3\arcsec$), decreasing to $\sigma_\mathrm{F336W-F343N} = 0.016$~mag at radii $R>70\arcsec$.

When using the artificial star tests to estimate the error distributions for observed stars in NGC~2419, we need to correct for the fact that the radial distribution of the artificial stars does not exactly match that of the actual cluster stars. To this end, we assigned a radius-dependent weight $w_A(R)$ to each artificial star, where $w_A(R)$ is given by the ratio of the number of actual stars at radius $R$ to the number of artificial stars at the same radius. We computed these weights in radial bins of 100 pixels (4\arcsec), and then assigned the corresponding $w_A(R)$ to all artificial stars within that radial bin. The weights ranged between $w_A(R)\approx0.15$ and $w_A(R)\approx0.35$. This relatively modest variation, combined with the weak radial dependence of the errors, means that the error distributions and corresponding estimates of the photometric errors change little when applying the weights.

   \begin{figure}
   \centering
   \includegraphics[width=\columnwidth]{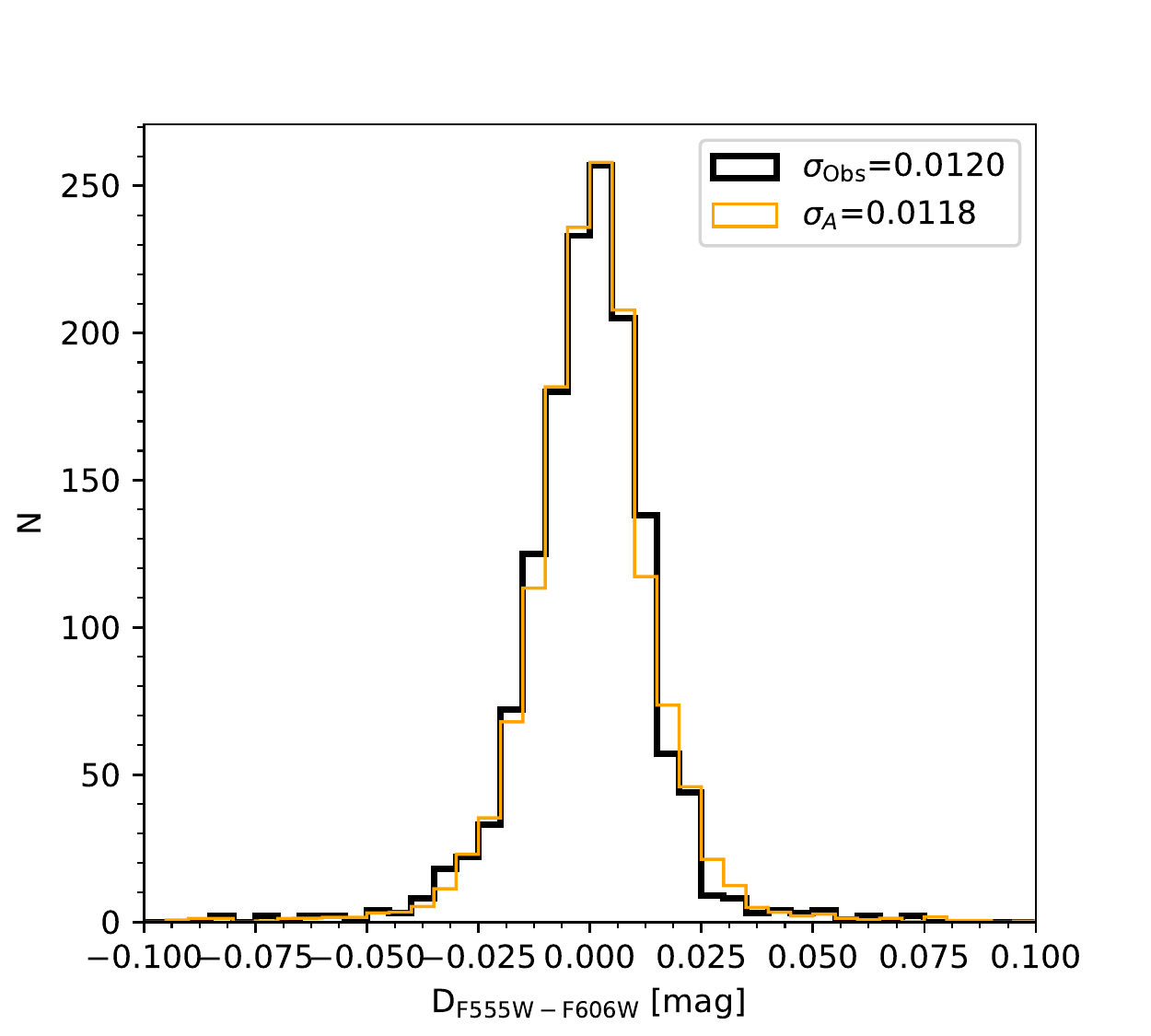}
      \caption{Histograms of the D$_\mathrm{F555W-F606W}$ distributions for observed RGB stars and artificial stars.
         \label{fig:dvr}
         }
   \end{figure}
   
While artificial star tests such as those described above are widely used, it should be kept in mind that it is nearly impossible to include all effects that are present in the real data in such tests \citep[see e.g.][]{Bellazzini2002,Anderson2008}. The PSF of the artificial stars will typically not be an exact match to that of the real stars, and fitting the artificial stars with the same PSF used to generate them would fail to account fully for uncertainties in the PSF modelling. While we have attempted to account for this by redetermining the PSF from artificial stars, cases where artificial stars are blended with real stars may still be problematic. Furthermore, the real PSF varies across the WFC3 detectors, whereas our artificial star tests used a single PSF for all stars.  Although spatial variations in the PSF are taken into account in the \texttt{ALLFRAME} photometry, the modelling of these variations is by necessity imperfect, and this is difficult to take into account in a fully realistic way in the artificial star tests. Even if the spatial variability of the PSF were fed back into the artificial star tests, it would be restricted to the parameterisation used by \texttt{ALLFRAME}.  

To assess the fidelity of the artificial star tests, Figure~\ref{fig:dvr} shows a comparison of the observed spread in the F555W-F606W colours around the RGB with the corresponding spread for the artificial stars.
The quantity D$_\mathrm{F555W-F606W}$ denotes the difference between the observed colours and a polynomial fit to the RGB for stars in the range $0 < M_\mathrm{F438W} < +2$. The F555W-F606W combination is expected to be insensitive to the abundance variations arising from the presence of multiple populations because of the small colour baseline and the lack of strong molecular features in these two bands, so that the spread in D$_\mathrm{F555W-F606W}$ should be due mainly to observational errors. As can be seen in the figure, the distributions for the observed and artificial stars are very similar indeed. To be more quantitative, 
we again estimated the dispersions from the 16th and 84th percentiles of the D$_\mathrm{F555W-F606W}$ colour distribution. For a Gaussian distribution, this will be similar to the dispersion computed in the usual way, but by using the percentiles we are less sensitive to extreme outliers. 
As indicated in the legend, the dispersion for the artificial stars ($\sigma_A = 0.0118$ mag) is very similar to that of the observations ($\sigma_\mathrm{Obs} = 0.0120$ mag). We note that including the correction for spatial coverage of the artificial stars makes virtually no difference; if this correction is omitted we find
 $\sigma_A = 0.0119$~mag.
 Very similar results were obtained from the F814W-F850LP combination, for which the same comparison yields $\sigma_A = 0.0153$~mag and $\sigma_\mathrm{Obs} = 0.0147$~mag.
 
We conclude that the artificial star tests, in spite of the potential concerns mentioned above, provide a fairly realistic estimate of the random uncertainties in our photometric analysis.

\section{Results}
   
   \begin{figure}
   \centering
   \includegraphics[width=\columnwidth]{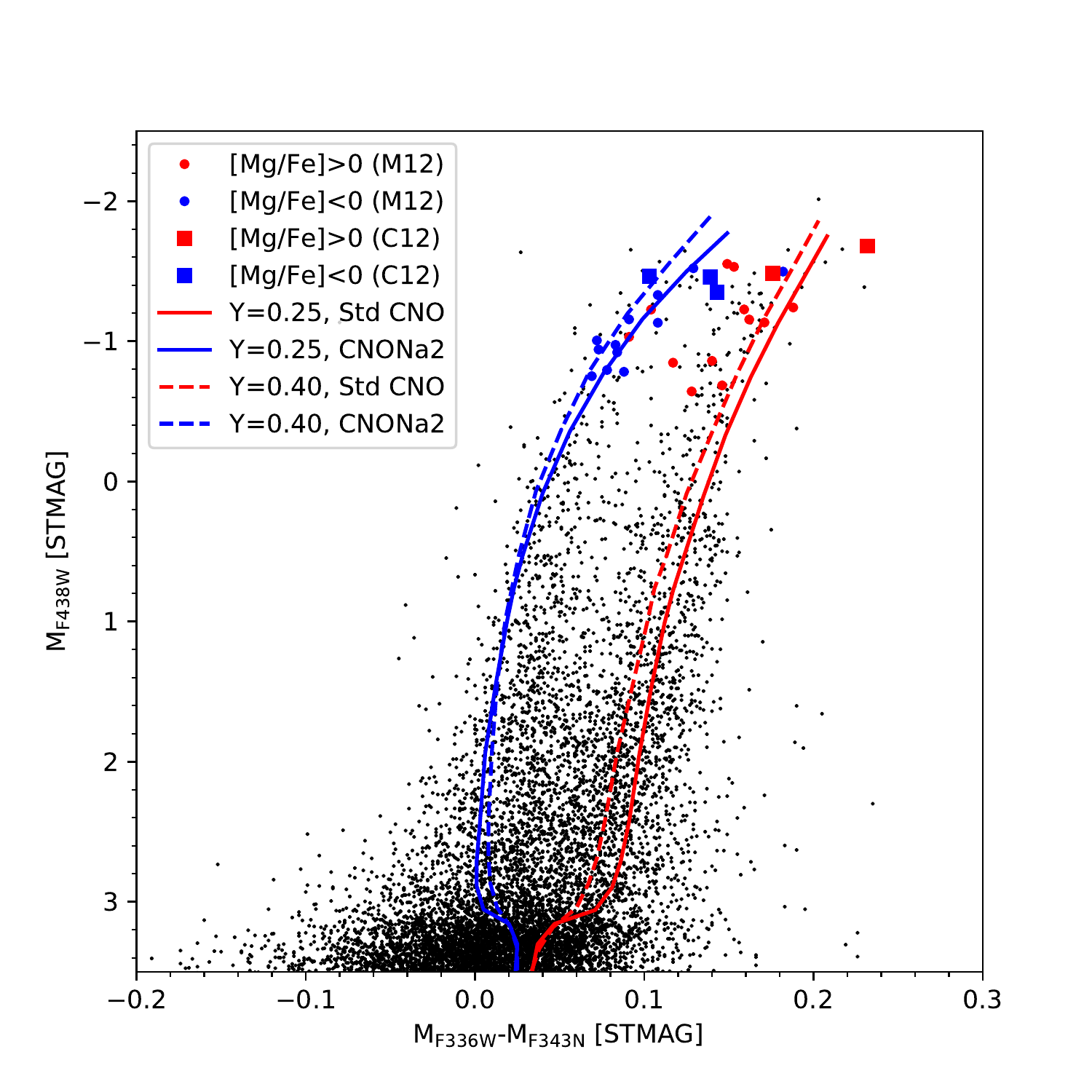}
      \caption{$M_\mathrm{F438W}$ vs.\ $M_\mathrm{336W}-M_\mathrm{F343N}$ diagram. The filled coloured symbols are stars with $\mathrm{[Mg/Fe]}$ measurements  \citep{Cohen2012,Mucciarelli2012a}.
         \label{fig:uuni}
         }
   \end{figure}

   \begin{figure}
   \centering
   \includegraphics[width=42mm]{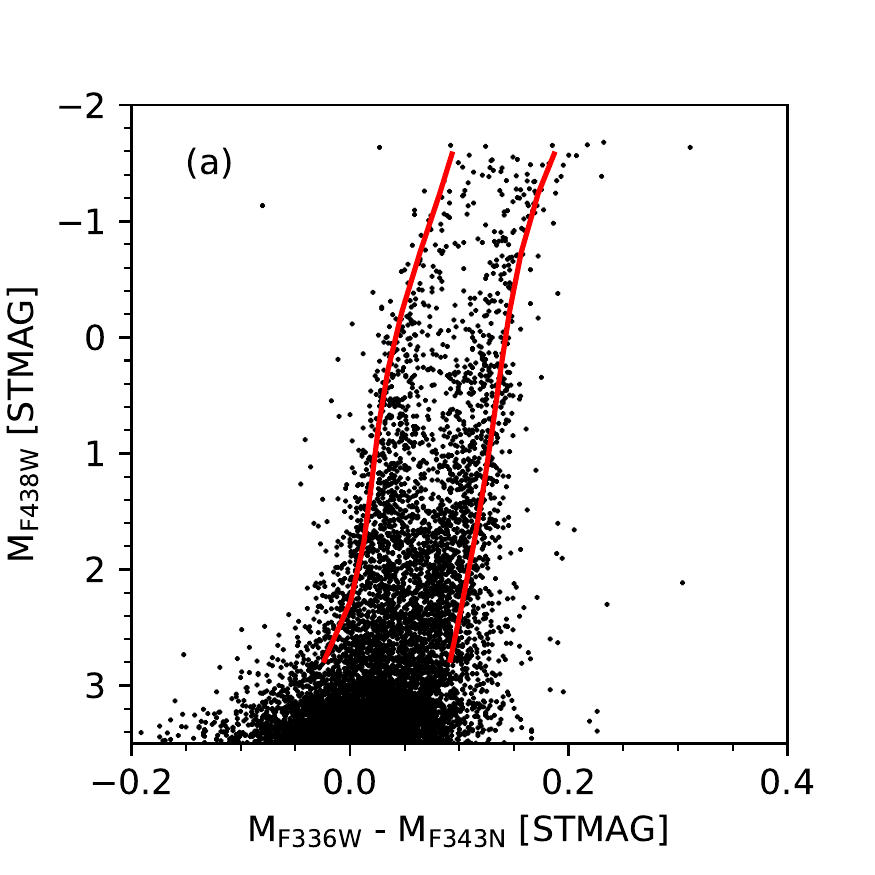}
   \includegraphics[width=42mm]{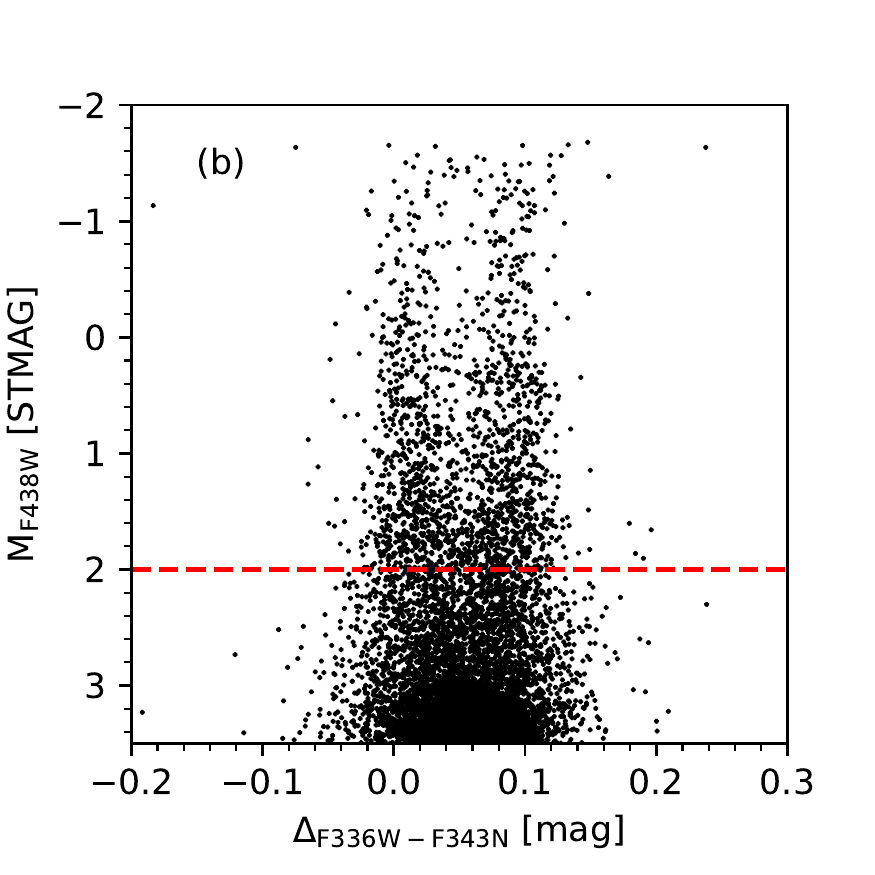}
      \caption{(a): F438W vs.\ F336W-F343N colour-magnitude diagram with the 10\% and 90\% fiducial lines indicated. 
         (b): verticalised CMD.
         \label{fig:cmd_unu_b}
         }
   \end{figure}

   \begin{figure}
   \centering
   \includegraphics[width=\columnwidth]{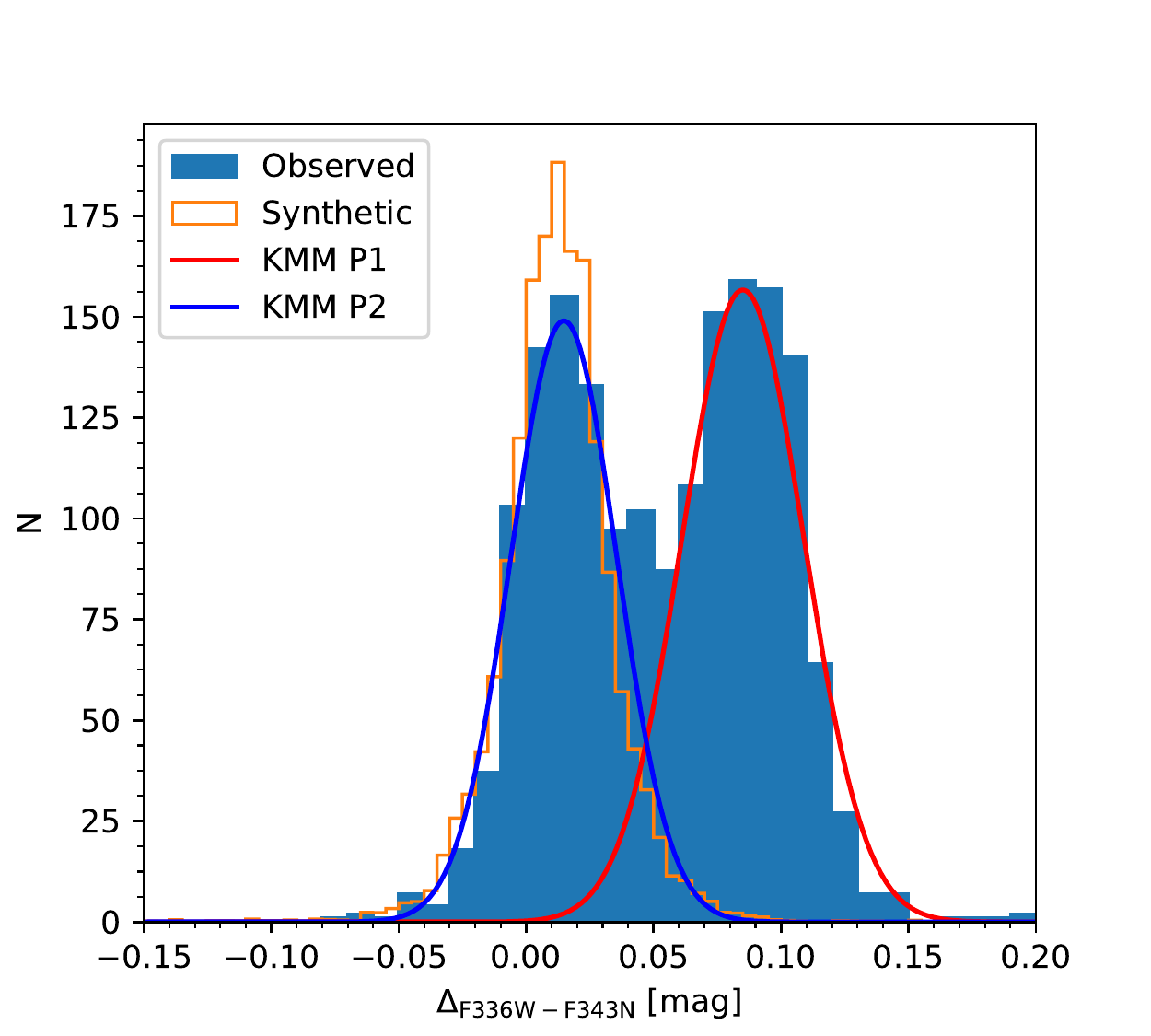}
      \caption{Histogram of the $\Delta_\mathrm{F336W-F343N}$ colour distribution. The error distribution of the synthetic stars has been shifted by 0.015 mag to match the P2 peak. 
         \label{fig:hist_dunu}
         }
   \end{figure}

\subsection{Ultraviolet colours and N abundance variations}
\label{sec:uvcol}

   In Fig.~\ref{fig:uuni} we show the $M_\mathrm{F438W}$ vs.\ $M_\mathrm{F336W}-M_\mathrm{F343N}$ diagram for RGB stars in NGC~2419. Here, and in the rest of the paper, we assume that the extinction in F343N is the same as in F336W.
   We use the F438W magnitude instead of F336W on the vertical axis, since the latter turns over somewhat below the tip of the RGB (i.e., the coolest and most luminous RGB stars are not the brightest in F336W). The loci of N-normal and N-rich stars are therefore not well separated at the bright end when plotting $M_\mathrm{F336W}$ vs.\ $M_\mathrm{F336W}-M_\mathrm{F343N}$. Filters centred at even longer wavelengths would potentially be even better than F438W, but here saturation becomes problematic. 
   
   The stars for which spectroscopic Mg abundance measurements are available in the literature \citep{Cohen2012,Mucciarelli2012a} are highlighted with red ($\mathrm{[Mg/Fe]}>0$) and blue ($\mathrm{[Mg/Fe]}<0$) symbols.
Also included in the figure are synthetic colours computed for standard field-like ($\alpha$-enhanced) and CNONa2 mixture, for He content of $Y=0.25$ (solid lines) and $Y=0.40$ (dashed lines). We used isochrones with $\mathrm{[Fe/H]}=-2$ and $t=13$ Gyr from the Dartmouth database \citep{Dotter2007} and colours were calculated with \texttt{ATLAS12}/\texttt{SYNTHE}, as described in Sect.~\ref{sec:artstar}. We also computed a set of model atmospheres and spectra in which the Mg abundance was decreased by 1 dex, but this was found to have a negligible effect on the colours.
The model colours have been shifted by a small offset (0.02 mag) in the horizontal direction.

It is clear that the F336W-F343N colour index is a very effective discriminator of CNO content, specifically N abundance, while it is hardly affected by He content. The observed spread in F336W-F343N is roughly similar to the separation between the field-like and CNONa2 models, and much larger than the photometric errors. Henceforth we refer to stars appearing to the right in Fig.~\ref{fig:uuni} (i.e., those with field-like composition) as belonging to P1 and those to the left as P2. We will adhere to this nomenclature specifically in the context of CNO variations. It is clear that the Mg-poor stars are associated mostly with the N-rich population P2, and the Mg-normal stars mostly with the N-normal population P1, although there may not be an exact 1:1 correspondence.

Figure~\ref{fig:uuni} already hints at a bimodal distribution of the F336W-F343N colours. To further examine the properties of the colour distribution, we verticalised Fig.~\ref{fig:uuni}, following a similar procedure to that described in \citet{Milone2017}.
This entailed computing the offset in F336W-F343N for each star with respect to a fiducial line, and renormalising the offsets relative to those at some fixed magnitude to account for changes in the width of the colour distribution. While the model lines in Fig.~\ref{fig:uuni} roughly trace the extremes of the F336W-F343N colour distribution, it was found that a better result was obtained by determining the fiducial lines directly from the data. We computed the 10\% and 90\% percentiles of the F336W-F343N colours as a function of $M_\mathrm{F438W}$ in bins of 0.5 mag, and then fitted fourth-order polynomials to the percentile values vs.\ $M_\mathrm{F438W}$. In reality, the 10\% and 90\% lines were found to be nearly parallel (Fig.~\ref{fig:cmd_unu_b}, panel (a)), with the separation varying between 0.101 mag at $M_\mathrm{F438W} = +2$ and 0.104 mag at $M_\mathrm{F438W} = +0.5$, and then narrowing slightly to 0.092 mag at $M_\mathrm{F438W} = -1.5$.
We then defined the offset $\Delta_\mathrm{F336W-F343N}$ with respect to the 10\% percentile line, scaled to the separation between the two lines at a reference magnitude of $M_\mathrm{F438W} = +1$. 

 Panel (b) of Fig.~\ref{fig:cmd_unu_b} shows the verticalised CMD, in which two vertical sequences are readily visible. For magnitudes fainter than $M_\mathrm{F438W} = +2$ the separation between the two sequences becomes less evident, presumably because of the increasing photometric errors (Fig.~\ref{fig:synt_dunu_b}). In the following, we therefore restrict the analysis to stars brighter than $M_\mathrm{F438W} = +2$, as indicated by the horizontal dashed line in Fig.~\ref{fig:cmd_unu_b}. This gives a sample of 1717 RGB stars. 

Figure~\ref{fig:hist_dunu} shows the distribution of $\Delta_\mathrm{F336W-F343N}$ for stars selected as described above. The histogram confirms that the distribution is clearly bimodal.
The orange open histogram shows the error distribution, $\delta_{o-i}$(F336W-F343N) for artificial stars in the same magnitude range as the data included in the figure. The $\delta_{o-i}$(F336W-F343N) values denote the differences between the input and recovered F336W-F343N colours, which were scaled by the same magnitude-dependent factor as the $\Delta_\mathrm{F336W-F343N}$ offsets in order to facilitate comparison with the observed $\Delta_\mathrm{F336W-F343N}$ distribution.
The error distribution has also been corrected for differences in the radial distributions of artificial and actual stars, using the weights $w_A(R)$ defined in Sect.~\ref{sec:artstar}.
The error histogram appears only slightly narrower than the two peaks in the observed colour distribution, which suggests that much of the broadening of the peaks may be due to photometric errors. 

To quantify the evidence for bimodality, we applied the KMM test \citep{Ashman1994} to the $\Delta_\mathrm{F336W-F343N}$ distribution. The KMM algorithm models the parent distribution of an observed sample as a sum of multiple Gaussians, and compares the likelihood  obtained from the best fitting multi-Gaussian model with that obtained from a single Gaussian. The improvement of the multi-Gaussian fit with respect to a single Gaussian is expressed as a $p$-value. Here we used the KMM algorithm to carry out a double-Gaussian fit to the $\Delta_\mathrm{F336W-F343N}$ distribution, allowing the peaks ($\mu_1$ and $\mu_2$) and dispersions ($\sigma_1$ and $\sigma_2$) of both Gaussians to vary. The red and blue curves in Fig.~\ref{fig:hist_dunu} represent the best-fitting double-Gaussian model estimated by the KMM algorithm. 
The KMM algorithm returned a $p$-value of $p<10^{-5}$, signifying that a double-Gaussian fit is a highly significant improvement over a single Gaussian (as was already evident from the histogram). The peak colours for P1 and P2 are $\mu_1 = 0.085$~mag and $\mu_2 = 0.015$~mag and the dispersions are $\sigma_1 = 0.024$~mag and $\sigma_2 = 0.021$~mag. The KMM algorithm assigned 939 stars (55\%) and 778 stars (45\%) to P1 and P2, respectively. 

From the artificial star tests we find a dispersion of $\sigma_\mathrm{art} = 0.022$ mag based on the  scaled $\delta_{o-i}$(F336W-F343N) values.
This value is essentially independent of whether or not the weights $w_A(R)$ are applied ($\sigma_\mathrm{art} = 0.0227$ without the weights, 0.0224 when including them). This is very similar to the width of the P2 peak found by the KMM algorithm, and only slightly narrower than the P1 peak. Again, this is consistent with the visual impression from Fig.~\ref{fig:hist_dunu}, which shows the $\delta_{o-i}$(F336W-F343N) histogram to be very similar to the P2 Gaussian.

The KMM algorithm assigns stars with $\Delta_\mathrm{F336W-F343N}<0.047$ mag to P2 and those with $\Delta_\mathrm{F336W-F343N}>0.047$ mag to P1. In the remainder of this paper we associate stars with P1 and P2 based on this limit.   

\subsection{Optical colours and pseudo-chromosome maps}

   \begin{figure}
   \centering
   \includegraphics[width=\columnwidth]{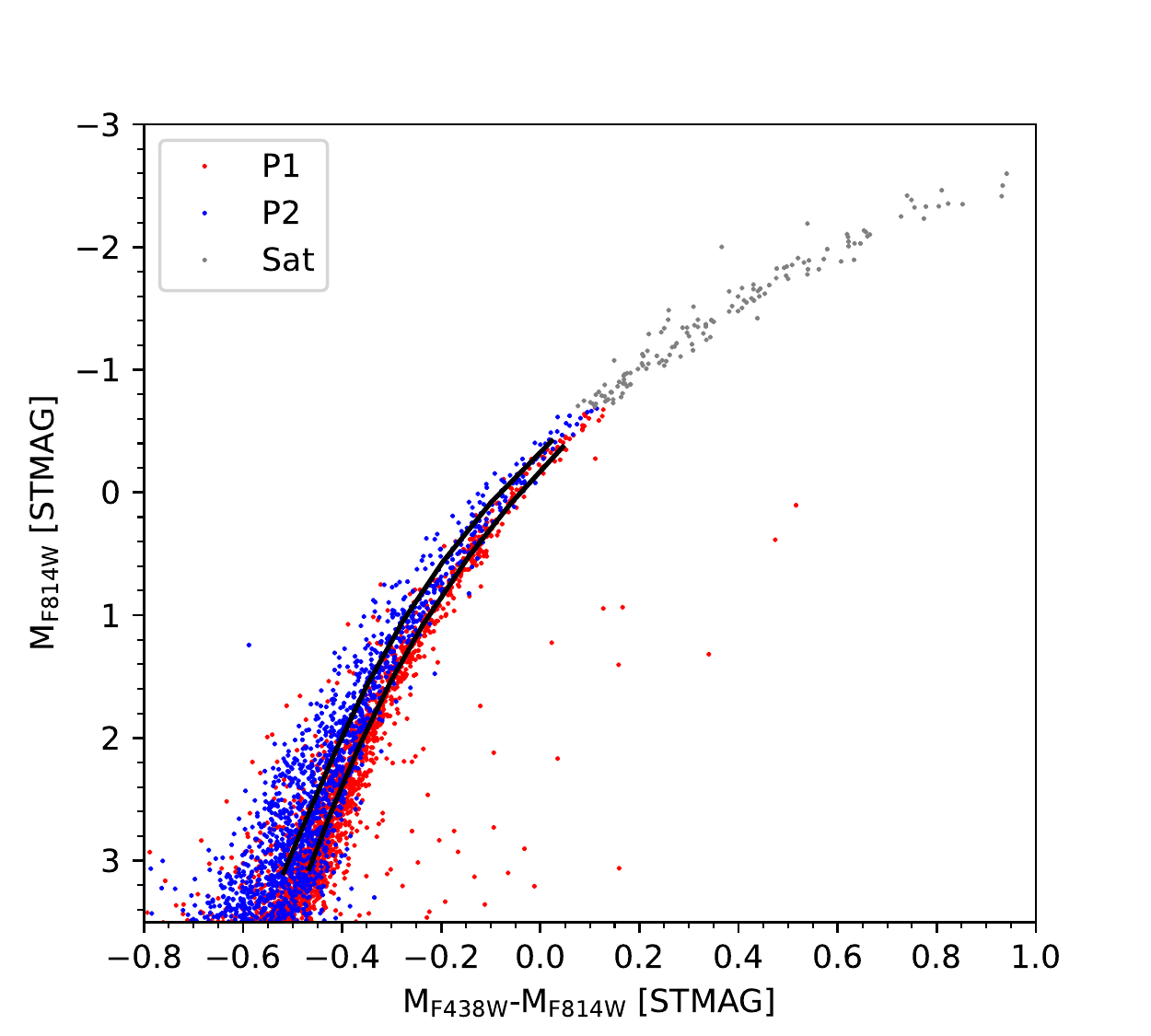}
      \caption{F814W vs.\ F438W-F814W  colour-magnitude diagram, showing the RGB for stars with $\Delta_\mathrm{F336W-F343N}>0.047$ (P1) and $\Delta_\mathrm{F336W-F343N}\le0.047$ (P2). The black curves are polynomial fits to the two sequences. Grey symbols indicate stars which are saturated in F814W.
         \label{fig:rgb_bi_i}
         }
   \end{figure}

   \begin{figure}
   \centering
   \includegraphics[width=\columnwidth]{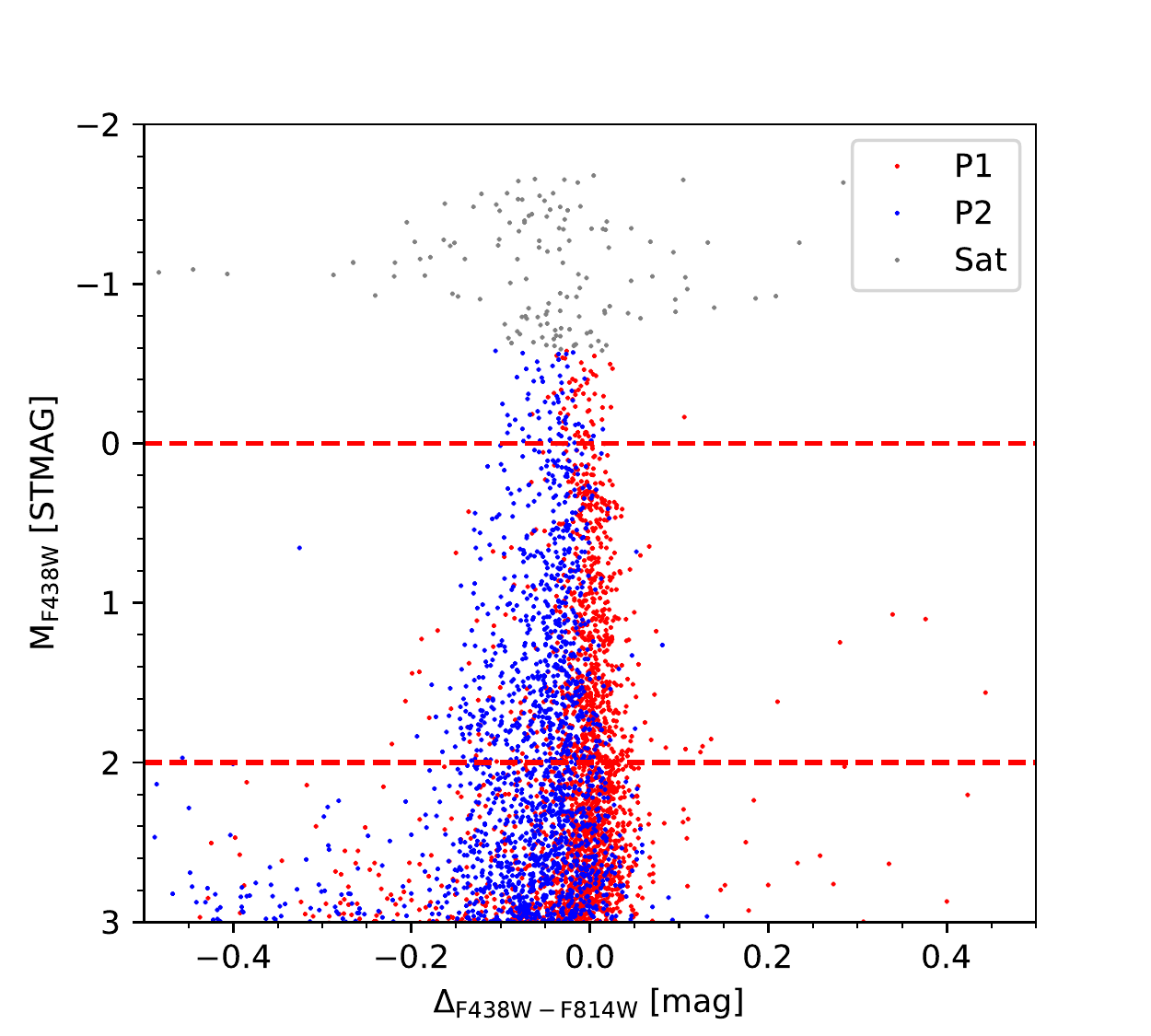}
      \caption{Verticalised F438W vs.\ F438W-F814W colour-magnitude diagram. The horizontal dashed lines show the magnitude limits used in the construction of the pseudo-chromosome map. Grey symbols indicate stars which are saturated in F814W.
         \label{fig:dbi_b}
         }
   \end{figure}

   \begin{figure}
   \centering
   \includegraphics[width=\columnwidth]{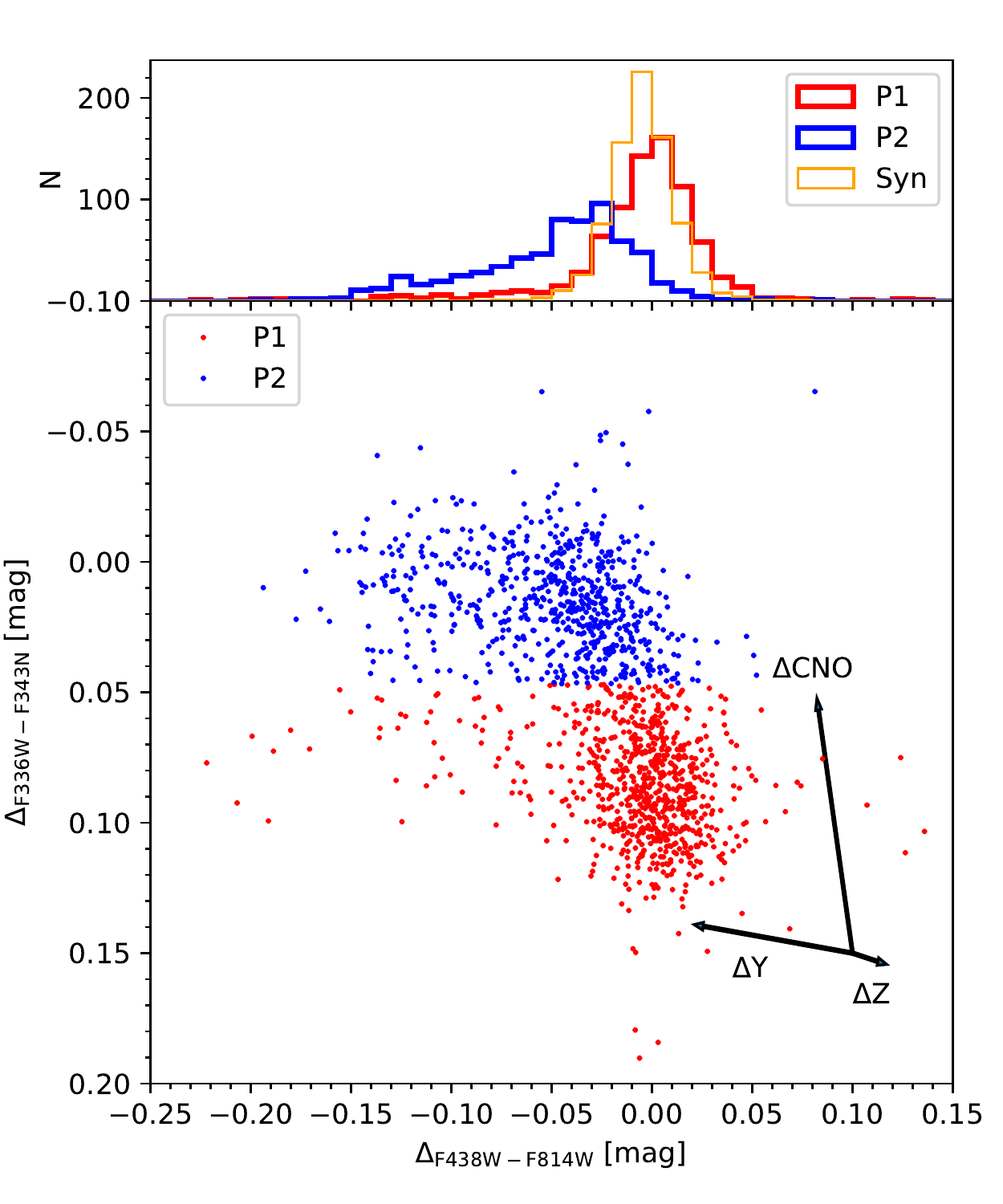}
      \caption{Pseudo-chromosome map and histograms of the $\Delta_\mathrm{F438W-F814W}$ colours for P1 and P2 stars.  The thin orange histogram indicates the errors as determined from artificial star tests. The arrows show the effect of changing the He abundance by $\Delta \mathrm{Y}=0.15$, the overall metallicity by $\Delta \mathrm{[Z/H]} = +0.1$ dex, and other light element abundances by $\Delta$([(C, N, O, Na)/Fe]) = $-0.6, +1.44, -0.8, +0.8$ dex \citep{Sbordone2011}.
         \label{fig:chromo}
         }
   \end{figure}

The F336W-F343N colours provide an effective and clean way to assess nitrogen abundance variations. However, it is becoming increasingly clear that more than one parameter may be required to characterise the abundance variations in GCs, even in mono-metallic clusters. A clear demonstration is provided by the diversity observed among the chromosome maps of GCs \citep{Milone2017}, which provide evidence of He abundance variations even among P1 stars in some clusters \citep{Lardo2018,Milone2018}.

While the chromosome maps defined by \citet{Milone2017} make use of the F275W filter, we can use the colours available for NGC~2419 to construct pseudo-chromosome maps that have the same basic ability to separate variations in N- and He abundance. For the N abundance variations we continue to rely on the $\Delta_\mathrm{F336W-F343N}$ index. To trace He abundance variations, we can use optical colours, preferably with a long baseline. Here we opt for the F438W-F814W combination, noting that F438W-F850LP gives essentially identical results. 

Figure~\ref{fig:rgb_bi_i} shows the $M_\mathrm{F814W}$ vs.\ $M_\mathrm{F438W} - M_\mathrm{F814W}$ CMD for RGB stars in NGC~2419. Stars associated with P1 and P2 based on their $\Delta_\mathrm{F336W-F343N}$ colours are plotted with red and blue symbols, respectively. It is evident that the P2 stars tend to have bluer F438W-F814W colours on average than the P1 stars, which is consistent with a higher He content for the P2 stars. We note that a difference in overall metallicity could potentially produce a similar effect on the RGB colours, in which case the bluer colours of the P2 stars would imply lower metallicities.

The two curves in Fig.~\ref{fig:rgb_bi_i} are polynomial fits to the P1 and P2 sequences. To verticalise the   CMD we follow a slightly different procedure than for F336W-F343N, since the photometric errors contribute significantly to the scatter. Instead of using percentiles, we use the fits to the P1 and P2 sequences as fiducials, calculating the offsets $\Delta_\mathrm{F438W-F814W}$ with respect to the P1 fiducial line and normalising, as before, at $M_\mathrm{F438W} = +1$ ($M_\mathrm{F814W} = +1.3$). The resulting verticalised version of Fig.~\ref{fig:rgb_bi_i} is shown in Fig.~\ref{fig:dbi_b}. Because of the saturation in F814W at bright magnitudes, we impose a bright magnitude limit of $M_\mathrm{F438W} = 0$, which leaves us with 1449 RGB stars. 

Figure~\ref{fig:chromo} shows the pseudo-chromosome map obtained by plotting $\Delta_\mathrm{F336W-F343N}$ vs.\ $\Delta_\mathrm{F438W-F814W}$. The arrows indicate the effect of changing the He content by $\Delta \mathrm{Y} = 0.15$, the overall metallicity by $\Delta \mathrm{[Z/H]} = +0.1$~dex, and the CNO abundances according to the CNONa2 mixture. We have defined the axes such that the plot resembles the chromosome maps of \citet{Milone2017} as closely as possible, i.e., with He content increasing towards the left and N increasing (and C and O decreasing) upwards. We note that the $\Delta$Z arrow is nearly anti-parallel to the $\Delta$Y arrow. 

The mean $\Delta_\mathrm{F438W-F814W}$ colours of the P1 and P2 stars are $-0.007$ mag and $-0.050$ mag, respectively.  Part of the colour difference may be caused by the differences in CNO abundances, since the $\Delta$CNO arrow is not exactly vertical. We will quantify this further below.  The histograms in the upper panel show the colour distributions of the P1 and P2 stars together with the distribution of $\delta_{o-i}$(F438W-F814W) from the artificial star tests. As was done for $\delta_{o-i}$(F336W-F343N), a magnitude-dependent scaling was applied to the $\delta_{o-i}$(F438W-F814W) values for consistency with the verticalised colour distribution.
For P1, the dispersion (as estimated from the 16th and 84th percentiles) is $\sigma_\mathrm{F438W-F814W} = 0.023$ mag and for P2 it is $\sigma_\mathrm{F438W-F814W} = 0.040$ mag. 
The standard deviation of the synthetic star colours is $\sigma=0.015$ mag. 
For the P2 stars, the $\Delta_\mathrm{F438W-F814W}$ distribution is thus significantly broader than the error distribution, and it appears that there may be some spread in  $\Delta_\mathrm{F438W-F814W}$ also for the P1 stars, with a tail towards the blue that may indicate the presence of He-enriched stars.

\subsection{Constraints on N and He abundance variations}

\begin{table*}
\caption{Observed and modelled colour differences between P1 and P2 stars.}
\label{tab:coldist}
\centering
\begin{tabular}{lccccccc}
\hline\hline
 Colour & $\langle\Delta(\mathrm{P2-P1})\rangle_\mathrm{obs}$ & $C_\mathrm{CNO}$ & $C_\mathrm{He}$ & $C_\mathrm{Z}$ & \multicolumn{3}{c}{$\Delta$(P2-P1)$_\mathrm{fit}$}
 \\ 
  & & & & & CNO, He & CNO, Z & CNO, He, Z \\ 
  & (mag) & & & & (mag) & (mag) & (mag) \\ 
  \hline
\hline
$\Delta_\mathrm{F336W-F343N}$ & $-0.070$ & $-0.093$ & $-0.010$ & $+0.003$ & $-0.081$ & $-0.087$ & $-0.071$ \\
$\Delta_\mathrm{F343N-F438W}$ & $+0.113$ & $+0.150$ & $-0.032$ & $+0.019$ & $+0.114$ & $+0.114$ & $+0.112$ \\
$\Delta_\mathrm{F336W-F438W}$ & $+0.045$ & $+0.056$ & $-0.043$ & $+0.022$ & $+0.032$ & $+0.026$ & $+0.039$ \\
$\Delta_\mathrm{F336W-F555W}$ & $+0.013$ & $+0.042$ & $-0.075$ & $+0.029$ & $+0.009$ & $+0.006$ & $+0.017$ \\
$\Delta_\mathrm{F336W-F814W}$ & $-0.001$ & $+0.042$ & $-0.105$ & $+0.032$ & $-0.001$ & $+0.003$ & $+0.001$ \\
$\Delta_\mathrm{F438W-F555W}$ & $-0.026$ & $-0.017$ & $-0.039$ & $+0.009$ & $-0.027$ & $-0.025$ & $-0.027$ \\
$\Delta_\mathrm{F438W-F814W}$ & $-0.044$ & $-0.017$ & $-0.075$ & $+0.013$ & $-0.040$ & $-0.029$ & $-0.045$ \\
$\Delta_\mathrm{F438W-F850LP}$ & $-0.048$ & $-0.017$ & $-0.081$ & $+0.013$ & $-0.042$ & $-0.029$ & $-0.049$ \\
$\Delta_\mathrm{F555W-F814W}$ & $-0.019$ & $+0.000$ & $-0.038$ & $+0.004$ & $-0.013$ & $-0.004$ & $-0.020$ \\
\hline
\end{tabular}
\tablefoot{The last three columns give the fitted colour differences $\Delta$(P2-P1)$_\mathrm{fit}$ when allowing for variations in CNO$+$He, CNO$+$Z, and CNO$+$He$+$Z, respectively.
}
\end{table*}

\begin{table}
\caption{Scaling coefficients for the fits to the colour differences in Table~\ref{tab:coldist}.}
\label{tab:coeff}
\centering
\begin{tabular}{lccc} \hline \hline
  & CNO, He & CNO, Z & CNO, He, Z \\ \hline
$\mathscr{S}_\mathrm{CNO}$ & 0.83 & 0.90 & 0.74 \\
$\mathscr{S}_\mathrm{He}$ & 0.34 & \ldots & 0.64 \\
$\mathscr{S}_\mathrm{Z}$ & \ldots & $-1.08$ & 1.17 \\ 
r.m.s. (mag) & 0.0065 & 0.0127 & 0.0024 \\ 
\hline
\end{tabular}
\tablefoot{The last row gives the r.m.s. dispersion of the differences $\langle\Delta(\mathrm{P2-P1})\rangle_\mathrm{obs} - \Delta$(P2-P1)$_\mathrm{fit}$ from Table~\ref{tab:coldist}.
}
\end{table}

   \begin{figure}
   \centering
      \includegraphics[width=8cm]{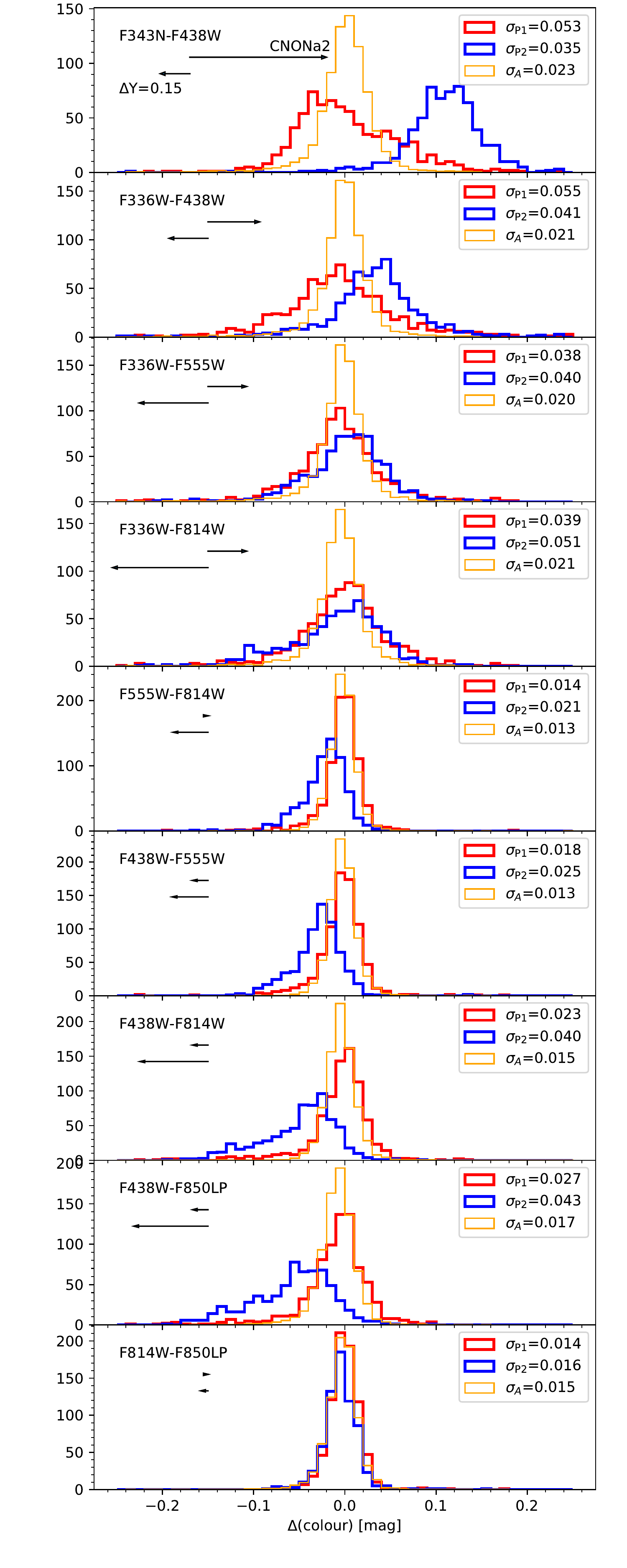}
      \caption{Colour distributions for P1 (red) and P2 stars (blue). The yellow histograms show the error distributions based on artificial star tests.
      The arrows indicate the effect of changing the He content by $\Delta$Y=$+0.15$ and the CNONa abundances by $\Delta$([(C, N, O, Na)/Fe]) = $-0.6, +1.44, -0.8, +0.8$ dex.
         \label{fig:hist_dcol}
         }
   
   \end{figure}

The pseudo-chromosome diagram utilises only two out of many possible colour combinations that can be formed from the available photometry. In Fig.~\ref{fig:hist_dcol} we show the histograms of various other colour combinations for the P1 and P2 stars, together with the error distributions from the artificial star tests. Again, the error distributions have been corrected for spatial coverage and the magnitude-dependent verticalisation scaling. The error distributions have been scaled to the same total number as the P1 stars.
Each panel also includes horizontal arrows indicating the effect of a change of 0.15 in the He abundance and a change in CNONa abundances from standard $\alpha$-enhanced abundances to the CNONa2 mixture. 

In many of these colour combinations, the two populations appear separated to some degree. For colour combinations that use F438W as the blue filter instead of F336W or F343N, the P2 stars generally appear bluer than the P1 stars, consistent with the P2 stars being more He rich (or metal-poor) on average. In these cases, the He and CNO arrows are parallel (due to the effect of C variations on the G-band contained within the F438W filter), so that He and CNO abundance variations reinforce each other.
For the combinations that do include F336W, the effect of modified CNO abundances tends to counterbalance the increase in He abundance for the P2 stars. For F336W-F814W these two effects appear to cancel out almost exactly on average, causing the colour distributions of P1 and P2 to be similar, whereas the smaller colour baseline for F336W-F555W implies that the CNONa variations win, making the P2 stars slightly redder on average than the P1 stars. This is reminiscent of the analysis by F2015, who found no difference between the Str{\"om}gren $u-y$ colours of Mg-poor and Mg-normal RGB stars in NGC~2419, and attributed this to the opposite effects of He and CNO abundance variations on this particular colour.
For F814W-F850LP (bottom panel), the two populations have essentially identical mean colours, which is consistent with the expectation that this colour combination should be insensitive to He- and CNO variations.

The dispersions of the P1, P2, and synthetic colour distributions are indicated in the legends of Fig.~\ref{fig:hist_dcol}. In all cases (except F814W-F850LP), the colour spread for the P2 stars clearly exceeds that expected from the artificial star tests. In general, the colour spread may be due to a combination of CNO and He abundance variations. The F555W-F814W colour is expected to be a relatively clean tracer of He abundance variations (as indicated by the short CNONa2 arrow) and the significant spread in F555W-F814W for the P2 stars therefore corroborates the conclusion from the chromosome map that a He spread is likely present within P2. For P1, the case for a significant internal He spread is less strong, since the observed F555W-F814W distribution is only slightly broader than the error distribution, although there is again a hint of an asymmetric tail towards the blue.
For F814W-F850LP we note that the observed dispersions of P1 and P2 are very similar to that of the artificial stars, which provides another verification that these tests give a realistic estimate of the photometric errors (cf.\ Sec.~\ref{sec:artstar}).

If we assume that the effects of He and CNO abundance variations, and potentially also of overall metallicity variations, can be combined linearly for each colour, then we can write
\begin{equation}
  \langle\Delta(\mathrm{P2-P1})_i \rangle = C_\mathrm{CNO,i}  \mathscr{S}_\mathrm{CNO}  + C_\mathrm{He,i}  \mathscr{S}_\mathrm{He}  + C_\mathrm{Z,i} \mathscr{S}_\mathrm{Z}
  \label{eq:dcol}
\end{equation}
where the coefficients $C_\mathrm{CNO,i}$, $C_\mathrm{He,i}$, and $C_\mathrm{Z,i}$ specify how the $i$th colour responds to variations in the CNO and He abundances and metallicity. Setting each of the scaling factors $\mathscr{S}_\mathrm{CNO}$, $\mathscr{S}_\mathrm{He}$, and $\mathscr{S}_\mathrm{Z}$ to unity corresponds to reference abundance changes of $\Delta[\mathrm{(C, N, O)/Fe}] = (-0.6, +1.44, -0.8)$~dex, $\Delta Y = 0.15$ (as indicated by the arrows in Fig.~\ref{fig:hist_dcol}) and $\Delta Z = 0.1$ dex. We can then solve for the scaling factors that need to be applied to the reference abundance changes  in order to best reproduce all of the observed colour differences $\langle\Delta(\mathrm{P2-P1})_i \rangle$.

In Table~\ref{tab:coldist} we list the mean observed colour differences between the P1 and P2 stars for each colour combination, with the exception of F814W-F850LP which contains little information and was included in Fig.~\ref{fig:hist_dcol} only as a consistency check.
The coefficients $C_\mathrm{CNO}$, $C_\mathrm{He}$, and $C_\mathrm{Z}$,  were calculated from our synthetic photometry at a reference magnitude of $M_\mathrm{F438W} = +1$.  The last three columns give the colour differences obtained by solving for variations in CNO and He (with Z fixed), in CNO and Z (with He fixed), and in all three parameters. 
In Table~\ref{tab:coeff} we give the corresponding best-fitting scaling factors $\mathscr{S}_\mathrm{CNO}$, $\mathscr{S}_\mathrm{He}$, and $\mathscr{S}_\mathrm{Z}$, which were found from a least-squares fit with the \texttt{lstsq} function in the \texttt{scipy.linalg} package in \texttt{Python}.

It is clear that, in all cases, a significant difference in mean CNO content between P1 and P2 is required to explain the colour differences, although the scaling factor $\mathscr{S}_\mathrm{CNO}$ is slightly less than unity for all fits. Hence, the implied average CNO difference between P1 and P2 is slightly smaller than assumed in the CNONa2 mixture. Table~\ref{tab:coldist} shows that a combination of variations in CNO and He content reproduces most of the P2-P1 colour differences to within about 0.01 mag, with an r.m.s.\ difference between the observed and modelled colour differences of 0.007 mag (Table~\ref{tab:coeff}). Assuming that $\Delta$Y scales linearly with $\mathscr{S}_\mathrm{He}$, the implied mean difference in He content is $\Delta \mathrm{Y} \simeq 0.05$.
Keeping He fixed and allowing Z to vary instead produces a somewhat worse fit, with an r.m.s.\ difference of 0.013 mag between the observed and best-fit colours. In this case, the metallicity scaling factor is negative, $\mathscr{S}_\mathrm{Z}=-1.1$, implying that P2 would be on average 0.11 dex more metal-poor than P1. Allowing all three scaling factors to vary produces the smallest residuals (r.m.s.\ = 0.002 mag), with a larger variation in He abundance ($\Delta \mathrm{Y} \simeq 0.10$) and P2 now being more metal-rich than P1 by about 0.12 dex. 

From the above, we conclude that small variations in mean metallicity may contribute to the observed colour differences, but these are largely degenerate with variations in He content and thus essentially unconstrained by our data. We can state that any differences in mean metallicity between P1 and P2 are likely less than about 0.1 dex, which is consistent with previous studies.
The variations in CNO are slightly smaller than those corresponding to the CNONa2 mixture, so we may estimate that the mean difference in N abundance between P2 and P1 is $\Delta \mathrm{[N/Fe]} \approx 0.9 \times 1.44~\mathrm{dex} \approx 1.3~\mathrm{dex}$. The mean difference in He content is $\Delta \mathrm{Y} \approx 0.05$, or possibly slightly larger if P2 is also more metal-rich. The model-dependent nature of these estimates should, however, be emphasised \citep[e.g.][]{Dotter2015}. 

\subsection{Radial distributions}
\label{sec:radist}

\begin{table}
\caption{Half-number radii for blue and red sub-populations, identified in various colours.}
\label{tab:rdist}
\centering
\begin{tabular}{l ccc}
\hline\hline
Colour & $R_h$(blue) & $R_h$(red) & $p_\mathrm{KS}$ \\ \hline
F336W-F343N & $31\farcs3\pm1\farcs1$ & $34\farcs6\pm0\farcs9$ & 0.046 \\
F343N-F438W & $35\farcs4\pm0\farcs8$ & $31\farcs4\pm1\farcs0$ & 0.020 \\
F336W-F438W & $33\farcs7\pm1\farcs1$ & $33\farcs5\pm1\farcs2$ & 0.708 \\
F336W-F555W & $33\farcs7\pm1\farcs2$ & $33\farcs5\pm1\farcs0$ & 0.247 \\
F336W-F814W & $32\farcs6\pm1\farcs2$ & $34\farcs0\pm1\farcs0$ & 0.051 \\
F438W-F555W & $31\farcs8\pm1\farcs1$ & $34\farcs8\pm0\farcs9$ & 0.100 \\
F438W-F814W & $31\farcs8\pm1\farcs2$ & $34\farcs8\pm0\farcs9$ & 0.050 \\
F438W-F850LP & $31\farcs8\pm1\farcs1$ & $35\farcs0\pm0\farcs9$ & 0.038 \\
F555W-F814W & $31\farcs8\pm1\farcs1$ & $35\farcs2\pm0\farcs9$ & 0.014 \\
F814W-F850LP & $33\farcs2\pm1\farcs2$ & $33\farcs7\pm1\farcs0$ & 0.350 \\
\hline
\end{tabular}
\tablefoot{$p_\mathrm{KS}$ gives the $p$-value from a Kolmogorov-Smirnov test when comparing the radial distributions of the blue and red sub-populations.}
\end{table}

   \begin{figure}
   \centering
   \includegraphics[width=\columnwidth]{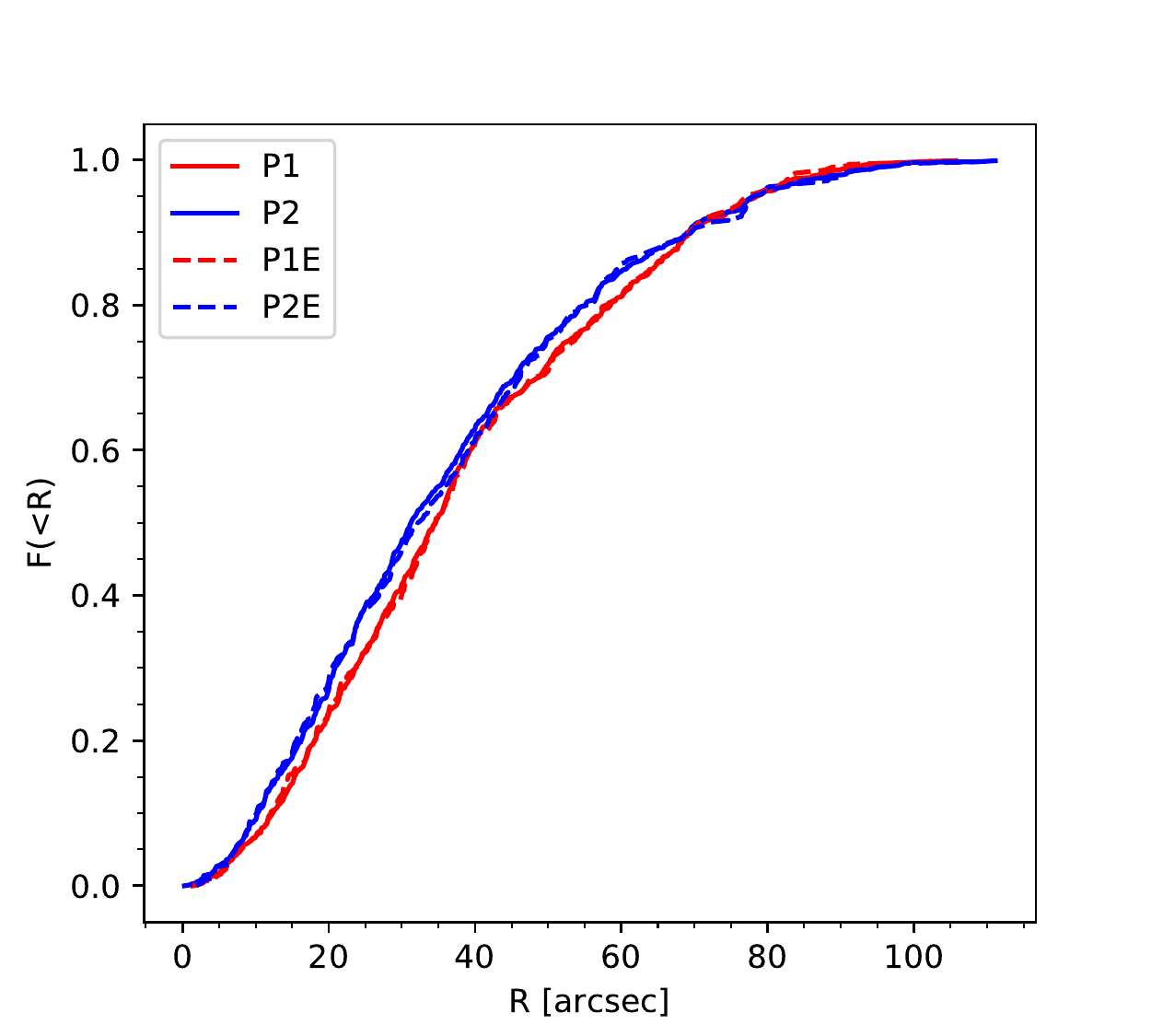}
      \caption{Cumulative radial distributions of P1 and P2 stars, selected based on their $\Delta_\mathrm{F336W-F343N}$ colours. P1E and P2E represent the ``extreme'' populations, which avoid the colour range between the peaks of P1 and P2.
         \label{fig:rdist_uun}
         }
   \end{figure}

As noted in the introduction, the long relaxation time of NGC~2419 means that the spatial distributions of sub-populations within the cluster are expected to be less affected by dynamical evolution than in other clusters. In this section we discuss the constraints on the radial distributions of subpopulations within NGC~2419.

In Fig.~\ref{fig:rdist_uun} we show the cumulative radial distributions of P1 and P2 stars.  The P2 stars are slightly more concentrated, but the difference between the P1 and P2 cumulative distributions does not appear dramatic and a Kolmogorov-Smirnov test returns $p_\mathrm{KS}=0.046$. From the cumulative distributions, the half-number radii for P1 and P2 are $34\farcs6\pm0\farcs9$ and $31\farcs3\pm1\farcs1$, respectively, with the $\sim10$\% difference being significant at about 2.3$\sigma$ (the errors were estimated via bootstrapping).
To get cleaner samples of P1 and P2 stars, we omitted stars with colours in the overlapping region between the peaks of P1 and P2, $\mu_2 < \Delta_\mathrm{F336W-F343N} < \mu_1$. Thus defining the extreme populations as P1E and P2E for $\Delta_\mathrm{F336W-F343N} > \mu_1$ and $\Delta_\mathrm{F336W-F343N} < \mu_2$, respectively, we get half-number radii of  $34\farcs6\pm1\farcs1$ (P1E) and $32\farcs2\pm1\farcs7$ (P2E). The corresponding cumulative distributions, which are shown with dashed lines in Fig.~\ref{fig:rdist_uun}, are very similar to those of the full P1 and P2 samples, but due to the smaller numbers of stars the Kolmogorov-Smirnov test now gives $p_\mathrm{KS}=0.30$, indicating no significant difference.
These half-number radii are all smaller than the half-light radii quoted in the literature, but this is as expected because the HST data do not include the outer parts of the cluster. 

In general, analyses of radial trends can be affected by variations in differential reddening as well as instrumental effects such as variations in the PSF and flat-field errors across the field-of-view. Because of the very similar central wavelengths of the F336W and F343N filters and the small foreground extinction towards NGC~2419, it appears unlikely that reddening variations could produce significant spurious trends in the $\Delta_\mathrm{F336W-F343N}$ index. 

To quantify the radial distributions in other colour combinations, we divided the RGB stars in NGC~2419 into blue and red samples in each colour, assigning (for simplicity) an equal number of stars to both groups.  
Table~\ref{tab:rdist} lists the half-number radii for the blue and red samples for each of the colour combinations shown in Fig.~\ref{fig:hist_dcol}. 
In most of these colour combinations, differences in the radial distributions are only marginally significant.
The most significant differences are found for F555W-F814W ($p_\mathrm{KS}=0.014$) and F343N-F438W ($p_\mathrm{KS}=0.020$).
For F343N-F438W, the top panel in Fig.~\ref{fig:hist_dcol} shows that this colour combination provides a fairly clear separation of the P1 and P2 stars, with the P2 stars having redder colours and being more concentrated. 
In the case of F555W-F814W, it is the blue stars that are more centrally concentrated. Blue F555W-F814W colours indicate enhanced He, which is characteristic of the P2 stars, so this is again consistent with the P2 stars being more centrally concentrated.
These results are difficult to attribute to differential reddening variations, for which we would expect the difference $R_h$(blue) - $R_h$(red) to always have the same sign.

Instrumental effects are harder to rule out definitively. 
One check is provided by the F814W-F850LP colour, which is expected to be insensitive to multiple populations. For this colour, Table~\ref{tab:rdist} shows no significant difference between the radial distributions of blue and red samples. This colour distribution is also the narrowest, and thus more liable to be affected by instrumental effects, so the fact that no differences between red and blue subsamples are seen in this colour combination supports the notion that the differences seen in other colours are real.

\subsection{Radial distributions: Helium or CNO as the main driver?}
\label{sec:hecno}

   \begin{figure}
   \centering
   \includegraphics[width=\columnwidth]{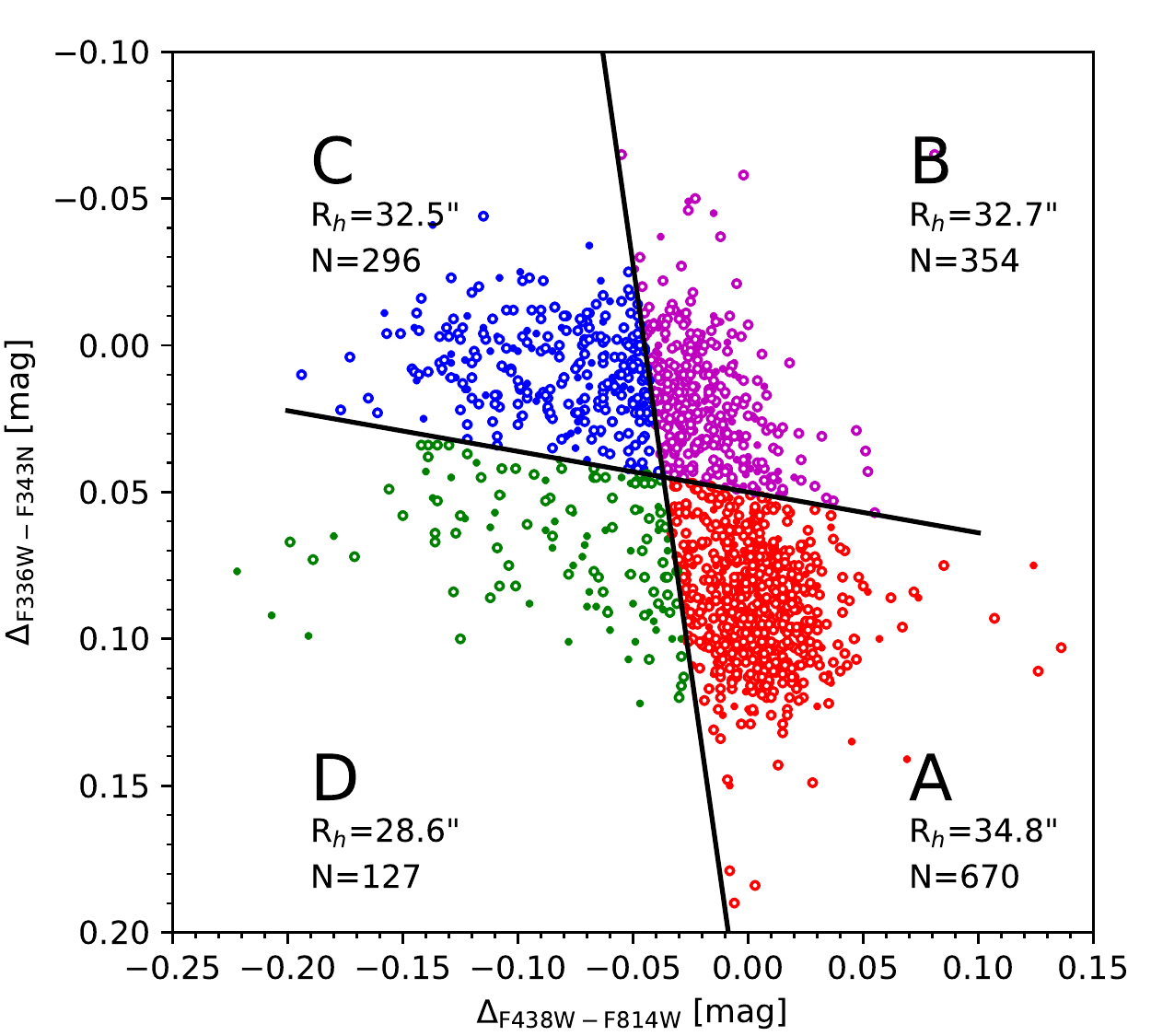}
      \caption{Chromosome diagram divided according to He and CNO content. The half-number radii are indicated in each quadrant. The high quality measurements ($\sigma_\mathrm{F438W} < 0.02$~mag, $\chi_\nu^2 < 2$) are shown with open circles.
         \label{fig:chromo_xy}
         }
   \end{figure}

   \begin{figure}
   \centering
   \includegraphics[width=\columnwidth]{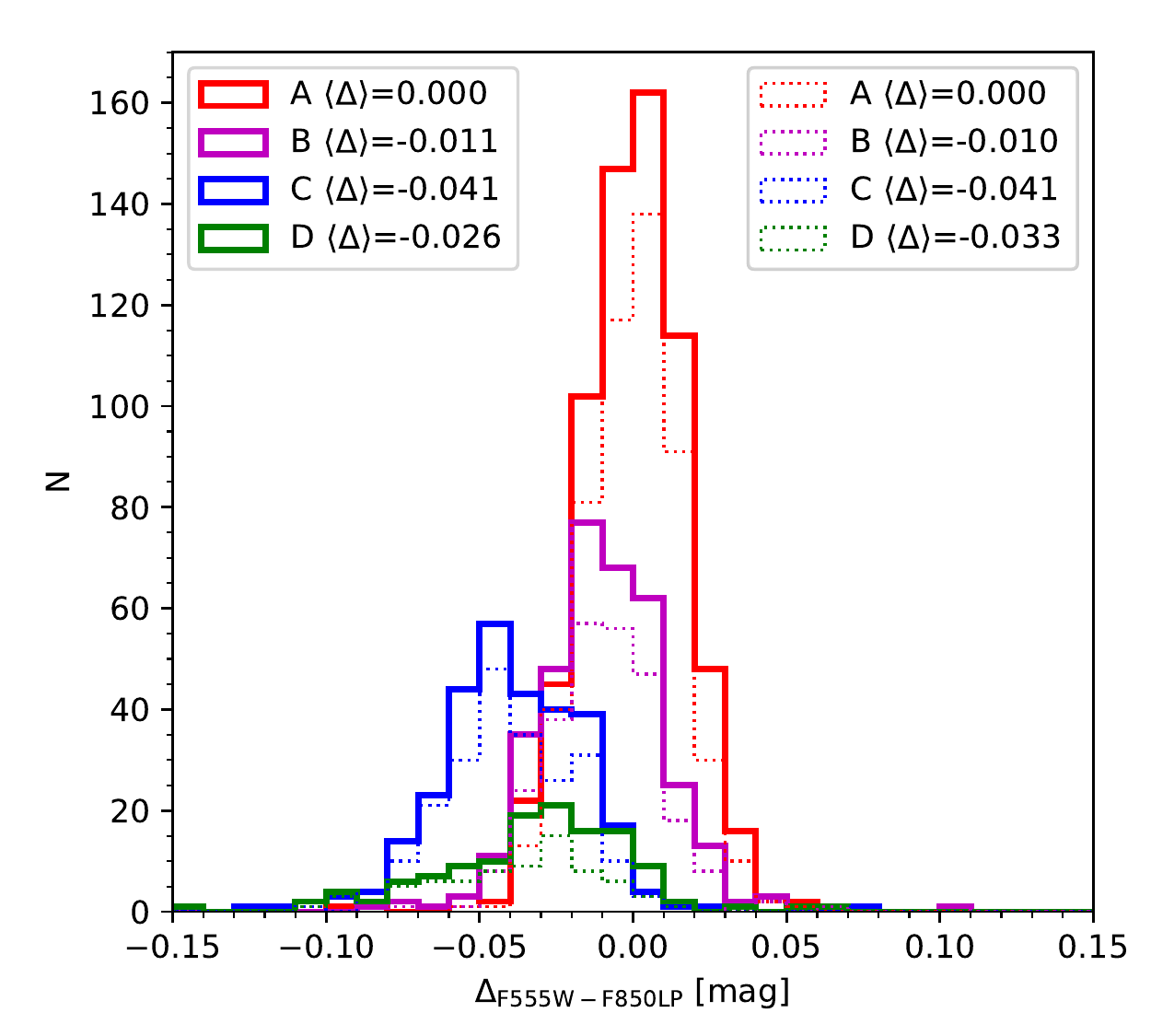}
      \caption{Colour distributions in $\Delta_\mathrm{F555W-F850LP}$ of the four regions identified in Fig.~\ref{fig:chromo_xy}. Histograms for the full samples are drawn with solid lines, whereas histograms for the stars remaining after error and $\chi^2$ cuts are shown with dotted lines.
         \label{fig:hist_dvl_abcd}
         }
   \end{figure}

   \begin{figure}
   \centering
   \includegraphics[width=\columnwidth]{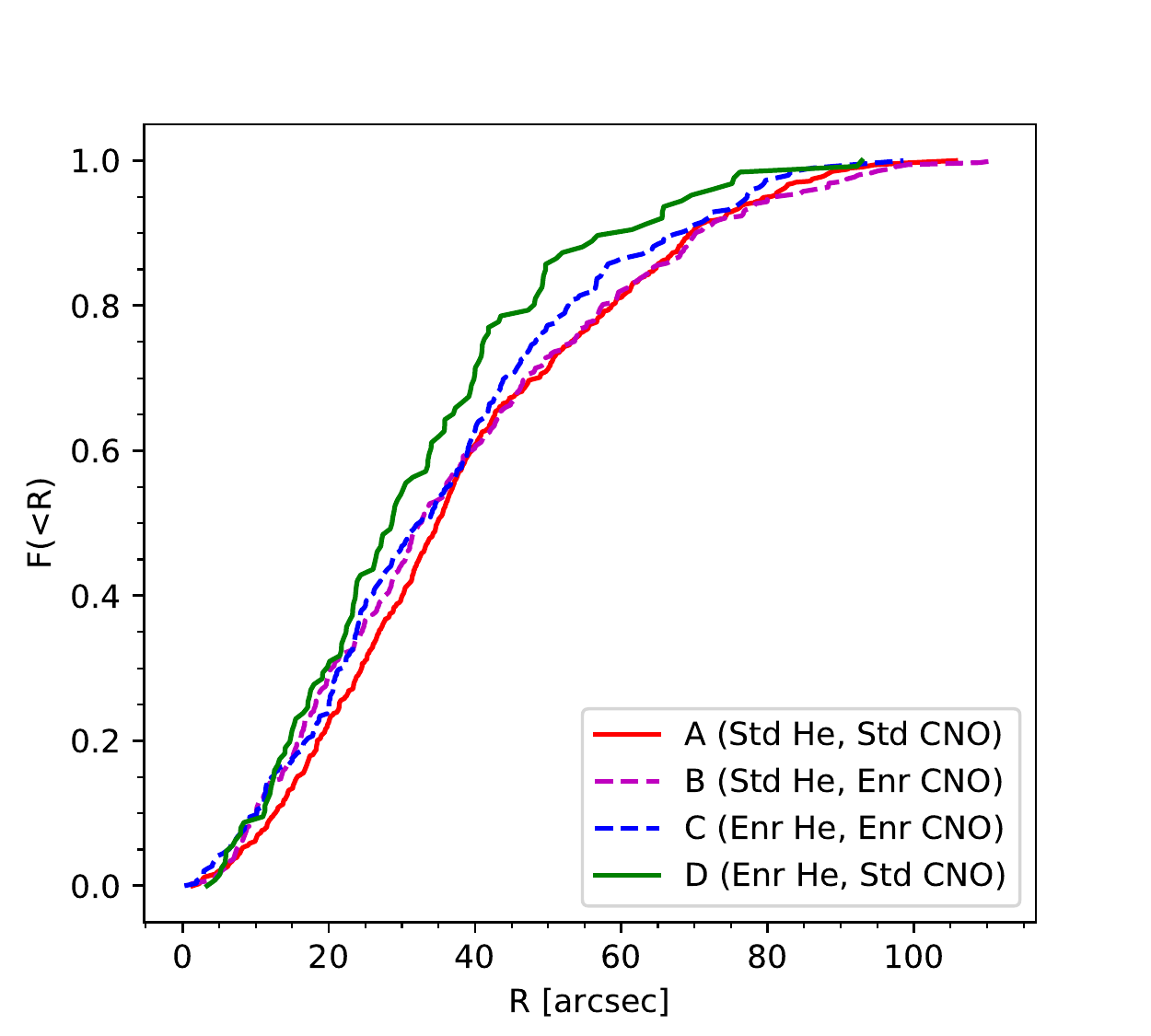}
      \caption{Cumulative radial distributions of stars selected on their He and CNO content. Colour coding is the same as in Fig.~\ref{fig:chromo_xy}.
         \label{fig:rdist_hecno}
         }
   \end{figure}

While the pseudo-chromosome diagram (Fig.~\ref{fig:chromo}) suggests some correlation between He and CNO abundance variations,  a significant He abundance spread is implied for at least the P2 stars, and possibly there is also some spread for the P1 stars. In the previous sections, we have used the CNO abundance variations as our primary means to split the cluster stars into sub-populations, with the tacit assumption that the intra-population He spreads can be treated as a perturbation on top of the CNO variations. This view seems to be supported by the clear bimodality in the CNO-sensitive colours, whereas the separation into distinct He-rich and He-normal groups is much less clear. 
Nevertheless, it appears worthwhile to examine in more detail how the radial distributions are affected by CNO as well as He abundance variations.

To this end, Fig.~\ref{fig:chromo_xy} again shows the pseudo-chromosome diagram, now divided into four regions. The stars with the highest quality photometry are shown with open circles (photometric errors in F438W less than $\sigma_\mathrm{F438W} = 0.02$~mag and \texttt{ALLFRAME} $\chi_\nu^2 < 2$) and other stars are shown with solid dots.
The black lines that separate the four regions have the same slopes as the arrows in Fig.~\ref{fig:chromo} and thus separate the stars by their He- and CNO content. 

Stars in the lower right-hand corner (region A) have the most field-like composition (normal He, normal CNO), whereas stars in the upper left-hand corner (region C) have the most GC-like composition (enhanced He and N). In each region we also indicate the half-number radius of the corresponding radial distribution and the number of stars. 
Fig.~\ref{fig:hist_dvl_abcd} shows the $\Delta_\mathrm{F555W-F850LP}$ distributions for the four regions defined in Fig.~\ref{fig:chromo_xy}. Histograms drawn with solid lines represent the full sample, and those drawn with dotted lines are for the high quality sub-sample.
Like $\Delta_\mathrm{F555W-F814W}$, the $\Delta_\mathrm{F555W-F850LP}$ combination is mainly sensitive to He abundance, and is used here because it is entirely independent of the colours used in Fig.~\ref{fig:chromo_xy}. The legend in Fig.~\ref{fig:hist_dvl_abcd} indicates the mean colour offsets relative to the A stars. These confirm that the C stars are the most He enriched population, and that the D stars also have a significant He enhancement relative to the A stars. Comparing the histograms for the full set of D stars with the high-quality subsample, it can be seen that the quality cut mainly removes stars with colours more similar to those of the A stars, which may have scattered into the D region. This leaves  a cleaner sample of pure D stars, so that the offset for the high-quality D stars is in fact slightly greater than for the full sample.
The offset of $-0.041$ mag between A and C corresponds to $\Delta\mathrm{Y} = 0.13$, which would suggest a He fraction as high as $Y\simeq0.38$ for the C stars. The C and D stars together correspond to about 29\% of the total number of RGB stars in the pseudo-chromosome map, which is comparable to the estimated fraction of RGB stars that are converted to extreme HB stars \citep{Sandquist2008}. It is thus tempting to associate the extreme HB stars with the He-enriched RGB stars in NGC~2419.

The cumulative radial distributions of the four populations are shown in Fig.~\ref{fig:rdist_hecno}. The least concentrated stars are those belonging to the A group, while the D stars are formally the most concentrated, but none of the differences are highly significant. Grouping the stars by He content, we find half-number radii of $34\farcs2$ for the (He-normal) AB stars and $30\farcs5$ for the (He-rich) CD stars, respectively, with a K-S $p-$value of $p_\mathrm{KS} = 0.03$ for the comparison of the cumulative distributions. Grouping instead by CNO content, the corresponding half-number radii are $34\farcs0$ (AD) and $32\farcs5$ (BC) with $p_\mathrm{KS}=0.25$. In both cases, the stars with the most field-like composition have the most extended distribution, and there is a suggestion that the correlation with He abundance may be somewhat stronger than with CNO content, in agreement with the results in Sec.~\ref{sec:radist}.

\subsection{Kinematics}

   \begin{figure}
   \centering
   \includegraphics[width=8cm]{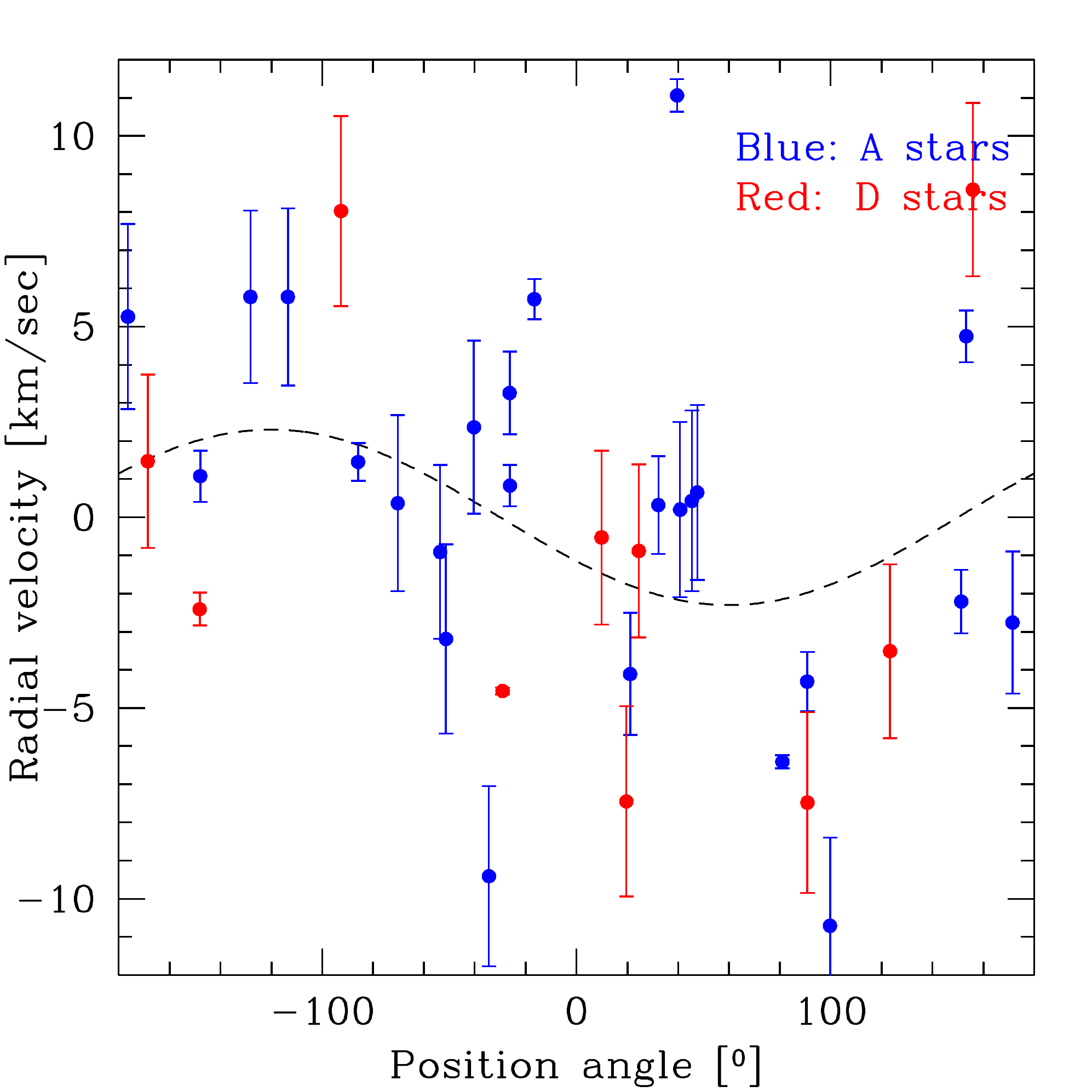}
   \includegraphics[width=8cm]{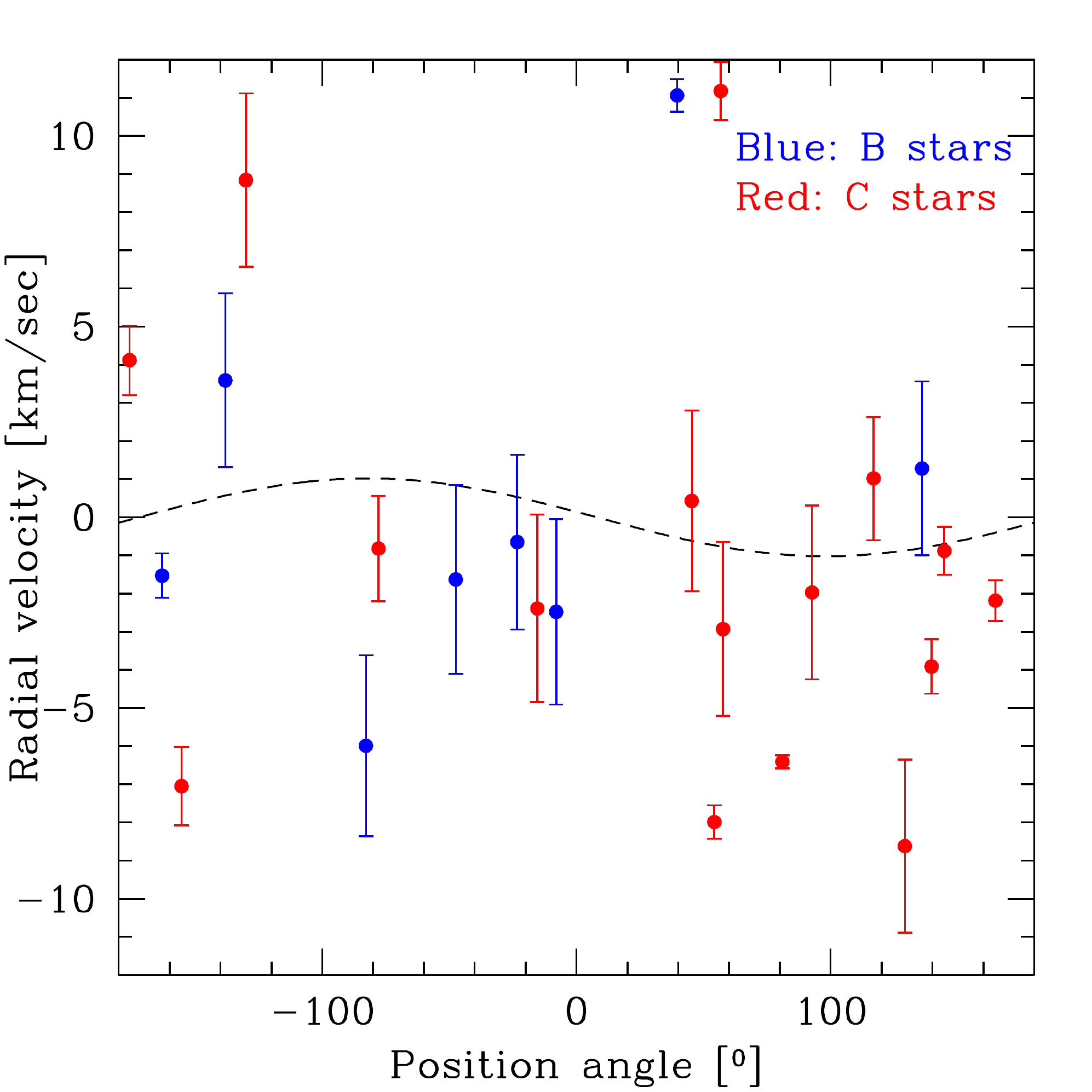}
      \caption{Kinematics. Top: AD (CNO-normal) stars. Bottom: BC (CNO-enriched) stars.
         \label{fig:kinematics}
         }
   \end{figure}

To test for kinematic differences between the different populations, we cross correlated the positions of the stars that we determined from the HST images against the positions of
stars with measured radial velocities from \cite{Baumgardt2018}, accepting stars with a positional difference of less than $0\farcs5$ as a match. This resulted in 64 matches.
We then subtracted the average velocity of NGC 2419, $v_r=-20.6 \pm 0.2$ km~s$^{-1}$ found by \cite{Baumgardt2018} from the velocities of all stars and applied a maximum-likelihood test according to
\begin{equation}
\ln \Lambda = \sum_i \frac{1}{\sqrt{ \sigma^2 + \epsilon^2_i}} e^{-\frac{1}{2}(v_i-v_{rot} \sin(\theta_i-\theta_0))^2/(\sigma^2 + \epsilon^2_i)}
\end{equation}
to determine the best-fitting values of the velocity dispersion $\sigma$, the amount of rotation $v_{rot}$ and the position angle $\theta_0$ of rotation for the different components. In the above formula, $\theta_i$ is the direction from the centre of NGC 2419 to a star (measured anti-clockwise from north), and $v_i$ and $\epsilon_i$ are the velocity and the velocity error of each star after subtraction of the mean velocity of NGC 2419. 

The stars with radial velocity measurements were assigned to P1 (standard CNO) and P2 (enriched CNO) using the $\Delta_\mathrm{F336W-F343N}$ index, as discussed in Sect.~\ref{sec:uvcol}. We obtained velocity dispersions of $\sigma = 4.76^{+0.66}_{-0.56}$ km~s$^{-1}$ for the 38 stars assigned to P1, and $\sigma = 5.78^{+0.89}_{-0.72}$ km~s$^{-1}$ for the remaining 26 P2 stars. 
Because these stars are mostly located within a few magnitudes of the tip of the RGB, assigning them to the He-normal and He-rich populations is more difficult. We made a rough assignment based on the F438W-F850LP colour. 
The corresponding dispersions for the standard and He-rich stars are $\sigma = 5.06^{+0.75}_{-0.62}$  km~s$^{-1}$ and $\sigma = 5.67^{+0.76}_{-0.62}$  km~s$^{-1}$ respectively. Hence there is some indication that the stars
with field-like abundances have lower velocity dispersions, in agreement with their more extended spatial distribution. However the differences between the populations are still within the error bars.
 
The only group of stars for which we find significant rotation are the P1 stars, for which we find a rotation velocity of $v_r=2.4 \pm 1.1$ km~s$^{-1}$ with a position angle of $\theta_0=337 \pm 25$ degrees. The CNO enriched (P2) stars, instead, do not show any significant rotation (Fig.~\ref{fig:kinematics}). 
We formally find a rotation velocity for the P2 stars of $1.02\pm1.47$ km~s$^{-1}$, which is compatible with the same rotation as the P1 stars, as well as no rotation.
These results could indicate that the P1 stars in NGC~2419 formed from an initially slowly rotating gas cloud which imprinted its rotation signature on these stars, while the P2 stars may have formed from kinematically more mixed gas.

To test the significance of the detection of rotation for the P1 stars, we randomly drew 38 velocities  from a Gaussian distribution with a dispersion of 5 km~s$^{-1}$ and errors randomly distributed
between 0.5 and 2.5 km~s$^{-1}$. We assigned random position angles to these velocities (i.e., no underlying rotation). We then measured how often a rotation signal with more than $2.2 \sigma$ was found. This happened in 43 out of 500 cases, corresponding to a false positive rate of about 8\%.

\section{Discussion}

Despite the unusual chemical abundance patterns found by spectroscopic studies, such as the presence of extremely Mg-depleted and K-enhanced stars \citep{Cohen2012,Mucciarelli2012a}, the photometric evidence indicates a relatively normal range of CNO abundance variations within NGC~2419. The N-sensitive $\Delta_\mathrm{F336W-F343N}$ index reveals a clearly bimodal distribution of N abundances, with a difference in average $\mathrm{[N/Fe]}$ value of $\sim1.3$ dex between N-normal P1 stars and N-rich P2 stars. 
Our analysis is consistent with that by \citet{Frank2015}, who found a range in $\mathrm{[N/Fe]}$ of less than 1.3~dex from Str{\"o}mgren photometry for the outer parts of the cluster. While it should be kept in mind that the photometric estimates of the $\mathrm{[N/Fe]}$ variations are uncertain, and the difference in average N abundance between P1 and P2 may underestimate the full range somewhat, the estimated N abundance variations in NGC~2419 are not particularly extreme compared to those found in other GCs, where a range up to $\sim2$~dex in $\mathrm{[N/Fe]}$ has been found \citep[e.g.][]{Yong2008}. 

Our photometry is consistent with previous evidence that a significant fraction of the stars in NGC~2419 have enhanced He abundances. Assuming that metallicity variations are negligible, we find a mean difference of $\Delta$Y$\simeq 0.05$ between P1 and P2, implying $\langle$Y$\rangle \simeq 0.30$ for the P2 stars if the P1 stars have a normal He fraction of Y$\simeq0.25$. However, the total range is likely greater, since both populations (especially P2) show evidence of an intrinsic He spread. 
NGC~2419 is yet another example of a cluster where significant He spreads are found within the sub-populations identified via CNO-sensitive colours \citep{Nardiello2018,Milone2017,Lardo2018,Milone2018}.
The difference between the mean He abundances of P1 and P2 is relatively large compared to those found by \citet{Milone2018}, who found mean differences between 0 and 0.05 in $\Delta$Y for Galactic GCs, but it is consistent with their result that the larger differences tend to be found in more massive clusters.
Again, it is worth emphasising that the absolute values of the He abundance spreads derived from the photometry should not be taken too literally. 

A double-Gaussian fit to the $\Delta_\mathrm{F336W-F343N}$ distribution assigns about 45\% of the RGB stars to the N-rich P2. For a massive GC like NGC~2419, this is a relatively low fraction of enriched stars, with enriched fractions of $\sim70$\% or more being common for clusters with masses approaching $10^6 M_\odot$ \citep{Milone2017,Bastian2018}. The enriched fraction estimated from a simple double-Gaussian fit would increase if we consider that some of the P1 stars may not be truly field-like, but the stars in region A of Fig.~\ref{fig:chromo_xy} still account for nearly 1/2 of all stars in the diagram.
These estimates refer to the inner regions of NGC~2419 covered by our HST photometry, and the global fraction of enriched stars could be even lower if there are radial gradients. According to \citet{Beccari2013}, the fraction of stars with blue $u\!-\!I$ and $u\!-\!V$ colours decreases significantly outwards for radii between 100\arcsec\ and 400\arcsec. If these blue colours indicate enhanced He abundance, as assumed by B2013, then this would indeed imply a decreasing enriched fraction outwards. However, the strong radial trend seen in the $u\!-\!V$ and $u\!-\!I$ colours is somewhat puzzling, given the very small separation between P1 and P2 in the (nearly) equivalent F336W-F555W and F336W-F814W colours, and the lack of strong radial trends in these colours in the inner regions of the cluster (Fig.~\ref{fig:hist_dcol} and Table~\ref{tab:rdist}).

   \begin{figure}
   \centering
   \includegraphics[width=\columnwidth]{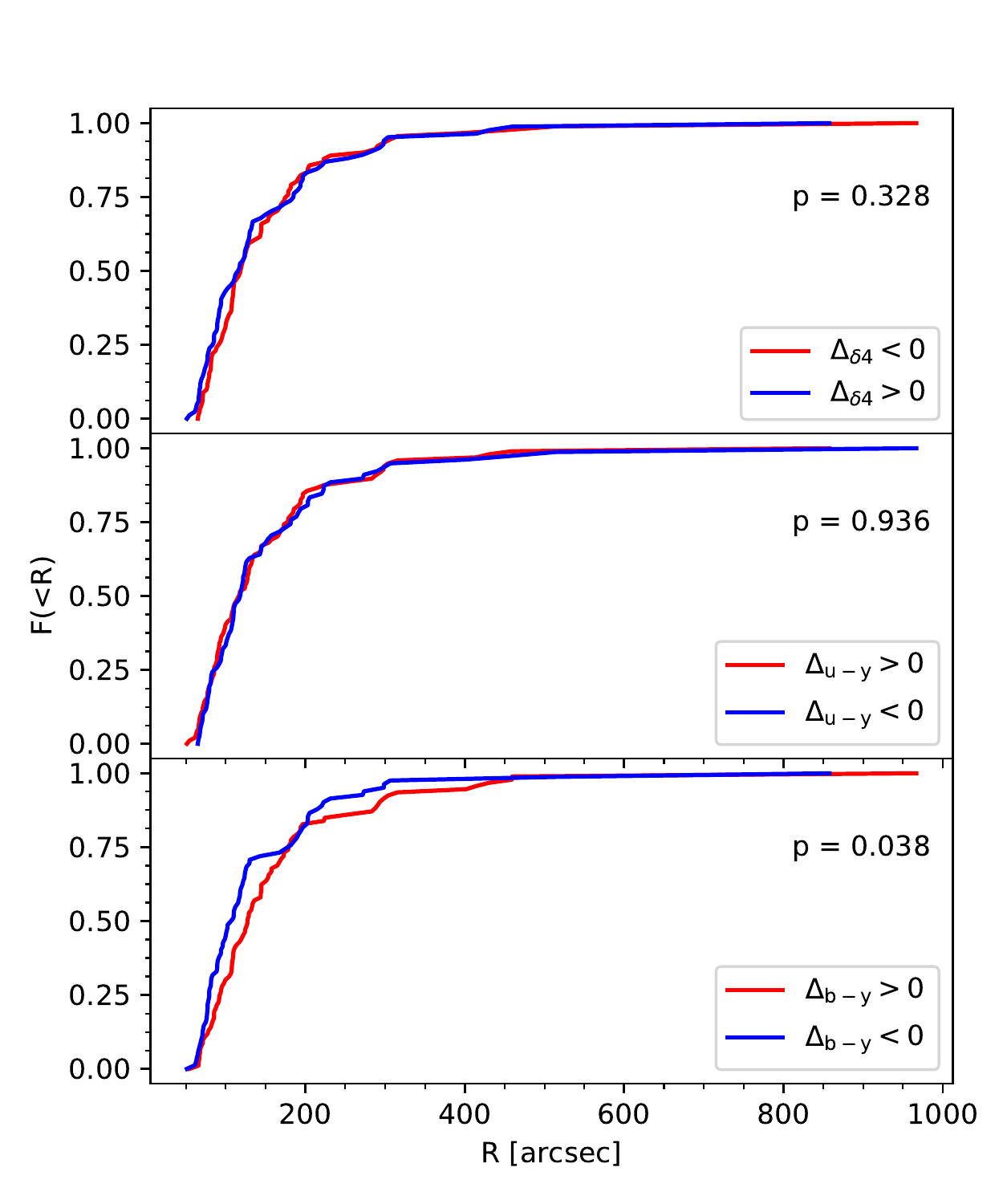}
      \caption{Cumulative radial distributions of RGB stars in the outer regions of NGC~2419, using Str{\"o}mgren photometry from \citet{Frank2015}
         \label{fig:frank}
         }
   \end{figure}

\citet{Frank2015} found an enriched fraction of $53\pm5$\% in the outer parts of NGC~2419 from a double-Gaussian fit to their measurements of the N-sensitive Str{\"o}mgren $\delta_4$ index. This is formally slightly higher than our estimate for the inner regions. F2015 did not investigate radial trends in detail, except by noting that their inferred enriched fraction is higher than that found for the inner regions of the cluster by other studies. To see whether their photometry can constrain radial trends, we downloaded their photometric catalogue and defined the $\Delta_{\delta_4}$ and equivalent $\Delta_\mathrm{b-y}$ and $\Delta_\mathrm{u-y}$ parameters with respect to a ridge line for the `clean RGB sample' in the same way that F2015 did. 

The comparisons of the resulting cumulative radial distributions for sub-samples divided according to $\Delta_{\delta_4}$, $\Delta_\mathrm{u-y}$, and $\Delta_\mathrm{b-y}$ are shown in Fig.~\ref{fig:frank}. The F2015 sample of RGB stars is smaller than our HST sample for the central regions (177 stars), so the statistical significance of any results is inevitably lower. Nevertheless, there is no significant difference between the radial distributions of N-normal and N-rich subsamples as defined by $\Delta\delta_4$ ($p_\mathrm{KS}=0.33$). 
Likewise, when dividing the sample according to $\Delta_\mathrm{u-y}$, the two radial distributions are essentially identical ($p_\mathrm{KS}=0.94$). This is seemingly at odds with the strong trends found by B2013, since the Str{\"o}mgren $u\!-\!y$ index is expected to behave very similarly to the $u\!-\!V$ colour. Indeed, F2015 found that the $u\!-\!y$ colours showed no correlation with Mg abundance, which they attributed  to the opposite effects of CNO and He abundance variations on this colour, a similar conclusion to that reached from our analysis. 
When dividing according to $\Delta_\mathrm{b-y}$, there is a somewhat significant difference ($p_\mathrm{KS}=0.038$), with the blue stars being more centrally concentrated. While the Str{\"o}mgren $b-y$ colour is insensitive to CNO abundance variations \citep{Carretta2011}, it does depend on He abundance (through $T_\mathrm{eff}$), and the mild tendency for the stars with blue $\Delta_\mathrm{b-y}$ colours (i.e.\ enhanced He) to be more centrally concentrated would be consistent with our results for the central regions.

The spectroscopically identified Mg-poor stars appear to be associated primarily with P2, and the Mg-normal stars with P1. 
So far, we have not discussed Na, which is perhaps the best established spectroscopic tracer of multiple populations. The Na-O anticorrelation is present in nearly all GCs where it has been looked for \citep{Carretta2009}, and it is thus natural to inquire about the behaviour of Na in NGC~2419. Unfortunately, the picture remains unclear in this regard. While \citet{Cohen2012} found a significant spread in $\mathrm{[Na/Fe]}$ within the cluster ($\sim 1$~dex), they found no correlation between $\mathrm{[Na/Fe]}$ and $\mathrm{[Mg/Fe]}$, with a difference of only 0.04~dex between the mean Na abundances of (five) Mg-deficient and (eight) Mg-normal stars. Five of the stars measured by \citet{Cohen2012} fall within our HST data; two of these happen to have low Na abundances (S810 and S1166) and also have normal Mg abundances. These two stars have F336W-F343N colours consistent with normal N abundances. The other three (S1004, S1065, S1131) are Mg-deficient and Na-rich, and their F336W-F343N colours are relatively blue, indicating enhanced N abundances (Fig.~\ref{fig:uuni}). In this sense, the behaviour of these five stars is as expected, but the broader implications for the behaviour of Na remain unclear,  given that many of the Mg-normal stars measured by \citet{Cohen2012} are, in fact, Na-rich.

Our data do not provide strong constraints on metallicity variations except that they are small, with a mean metallicity difference of $<0.1$ dex between P2 and P1. This is in agreement with the findings by most previous studies \citep{Mucciarelli2012a,Frank2015}, although there are also claims of a larger metallicity spread \citep{Lee2013}.

\section{Summary and conclusions}

We have used new HST/WFC3 imaging in the F343N and F336W filters, combined with archival optical HST/WFC3 data, to study the multiple populations in the central regions of the remote globular cluster NGC~2419. The data are spatially complete within a radius of $R=70\arcsec$ (28 pc), or about 1.5 projected half-light radii. The combination of UV and optical filters allowed us to constrain variations in He and N abundances for red giants in the inner regions of the cluster. We combined the photometry with radial velocity measurements from the literature \citep{Baumgardt2018} to examine the kinematics of the different populations.
Our main findings are as follows:

\begin{itemize}
\item The F336W-F343N colour distribution is clearly bimodal, as confirmed by a KMM test ($p<10^{-5}$). A double-Gaussian fit assigns 55\% of the stars to a population with F336W-F343N colours indicative of field-like nitrogen abundances (P1), and the rest to a population with nitrogen-enhanced composition (P2).
\item From a comparison of the mean optical colours of P1 and P2 stars with model calculations, we estimate a mean difference in the He content of $\Delta \mathrm{Y} \simeq 0.05$ between the two populations. Small metallicity differences ($<0.1$~dex) could also contribute to the colour differences. 
\item For the P2 stars, the observed spread in optical colours such as F555W-F814W and F438W-F814W is greater than the observational uncertainties. This most likely indicates a He spread at least within P2, with some stars possibly having He content as high as $Y\simeq0.38$. Analysis of the pseudo-chromosome map suggests that a small fraction (about 16\%) of the P1 stars may also be significantly He-enriched.
\item The P2 stars are somewhat more centrally concentrated within the cluster than the P1 stars, with some hint that the difference is driven primarily by differences in mean He content. The difference in the half-number radii is, at any rate, modest (about 10\%) and only moderately significant.
\item The P1 stars have a slightly lower velocity dispersion than the P2 stars, although the difference is not statistically significant. Nevertheless, the difference is in agreement with the more extended spatial distribution of the P1 stars. We find evidence of rotation for the P1 stars, whereas the data for the P2 stars are consistent with no rotation, as well as the same rotation as the P1 stars.
\item Stars for which spectroscopic measurements indicate a significant Mg-deficiency ($\mathrm{[Mg/Fe]}<0$) are associated primarily with the nitrogen-rich population.
\end{itemize}

In terms of the main elements studied and discussed in this paper, the abundance patterns seen in NGC~2419 are relatively unsurprising. The P1 stars identified through their field-like N abundances also tend to have relatively field-like He and Mg abundances, while the N-enriched P2 stars tend to have enhanced He and (strongly) depleted Mg. Nevertheless, the correlations between N and He have real scatter, and the same may well be true for N vs.\ Mg. It is well to keep in mind, however, the apparent lack of any correlation between Na and Mg. Here we cannot directly address the relation between Na and N or He, and this certainly appears to be a problem worthy of further investigation. 

The failure of current scenarios for the formation of GCs to provide a satisfactory account of the observed properties of multiple populations in general is well documented \citep{Bastian2018}, as are the additional problems associated with the complex chemistry of NGC~2419 specifically \citep[e.g.][]{DiCriscienzo2011,Cohen2012,Mucciarelli2012a,Carretta2013}.  The relative similarity of the radial distributions of the different populations found here poses yet another potential complication for many scenarios.

The possibility that NGC~2419 may be the nucleus of a disrupted dwarf galaxy has been discussed by many authors. Recent proper motion measurements, combined with the radial velocity of NGC~2419, appear to be consistent with membership of the Sagittarius dwarf spheroidal galaxy \citep{Massari2017,Sohn2018}, but since Sagittarius already has a nucleus \citep{Bellazzini2008} this would argue against NGC~2419 also being a nucleus.
It is not clear, in any case, that this would help explain its peculiar abundance patterns, which are not observed in dwarf galaxies \citep{Salgado2019}. Other more plausible candidates for nuclei (such as $\omega$~Cen and M54) display a chemical inventory that is quite different from what is seen in NGC~2419, with significant metallicity spreads but no reported Mg-K anticorrelation \citep{Carretta2013}. While NGC~2419 appears to represent a relatively extreme manifestation of the multiple populations phenomenon, it should be noted that the GC NGC~2808 shares some of the features seen in NGC~2419, such as the Mg-K anticorrelation \citep{Mucciarelli2015}, although other details differ. 

Hence, it seems that NGC~2419 falls within the range of behaviours that a successful theory for GC formation must be able to explain.

\begin{acknowledgements}
We thank the anonymous referee for a careful reading of the manuscript and several helpful comments. 
This work has made use of data from the European Space Agency (ESA) mission
{\it Gaia} (\url{https://www.cosmos.esa.int/gaia}), processed by the {\it Gaia}
Data Processing and Analysis Consortium (DPAC,
\url{https://www.cosmos.esa.int/web/gaia/dpac/consortium}). Funding for the DPAC
has been provided by national institutions, in particular the institutions
participating in the {\it Gaia} Multilateral Agreement.
N.B. gratefully acknowledges financial support from the Royal Society (University Research Fellowship) and the European Research Council (ERC-CoG-646928-Multi-Pop).
J.B. acknowledges support for HST Program number GO-15078 from NASA through grant HST-GO-15078.02  from the Space Telescope Science Institute, which is operated by the Association of Universities for Research in Astronomy, Incorporated, under NASA contract NAS5-26555.
\end{acknowledgements}

\bibliographystyle{aa}
\bibliography{refs.bib}

\begin{appendix}
\section{Photometry}
\begin{table*}
\caption{Photometry}
\label{tab:photometry}
\centering
\resizebox{\textwidth}{!}{
\begin{tabular}{cccccccccccccccccccccc} \hline\hline
 ID  &    X    &    Y   &     RA    &       DEC   &       F336W & eF336W & rF336W  & F343N & eF343N & rF343N &  F438W & eF438W  & F555W & eF555W &  F606W & eF606W &  F814W & eF814W &  F850L &  eF850L & Det\\ 
 (1)  &  (2)  &    (3)   &    (4)    &      (5)    &       (6)  &  (7)  &  (8)  &   (9) &  (10) &  (11)  &  (12) &  (13)  &  (14)  & (15)  &  (16) &  (17)  &  (18) &  (19)  &  (20)  & (21) & (22) \\ \hline
    15 & 1590.418 &   51.371 &  114.561106645 &  38.865331511 &   26.195 &  0.099 &  0.292 &   25.953 &  0.102 &  0.509 &   25.707 &  0.129 &   25.761 &  0.075 &   25.713 &  0.047 &   26.280 &  0.074 &   26.503 &  0.177 &   1 1   \\
    18 & 1974.725 &   51.343 &  114.556669290 &  38.862841127 &   26.355 &  0.116 &  0.221 &   26.352 &  0.121 &  0.335 &   26.065 &  0.151 &   26.039 &  0.079 &   26.034 &  0.072 &   26.354 &  0.084 &   26.797 &  0.210 &   1 1   \\
    30 & 2662.512 &   51.353 &  114.548704108 &  38.858370732 &   25.328 &  0.065 &  0.164 &   25.226 &  0.061 &  0.151 &   25.103 &  0.076 &   25.119 &  0.062 &   25.216 &  0.039 &   25.587 &  0.047 &   26.036 &  0.138 &   1 1   \\
    54 & 1985.385 &   51.496 &  114.556544707 &  38.862773300 &   24.141 &  0.042 &  0.069 &   24.108 &  0.034 &  0.173 &   23.648 &  0.041 &   23.908 &  0.021 &   23.995 &  0.026 &   24.510 &  0.023 &   24.756 &  0.051 &   1 1   \\
    57 & 2229.516 &   52.196 &  114.553714424 &  38.861194524 &   21.528 &  0.013 &  0.017 &   21.509 &  0.010 &  0.024 &   20.668 &  0.026 &   20.554 &  0.020 &   20.558 &  0.019 &   20.826 &  0.015 &   21.019 &  0.010 &   1 1   \\
    59 & 2312.434 &   51.733 &  114.552758674 &  38.860651661 &   26.073 &  0.098 &  0.184 &   26.139 &  0.112 &  0.794 &   26.161 &  0.116 &   26.067 &  0.059 &   26.070 &  0.073 &   26.530 &  0.104 &   26.711 &  0.174 &   1 1   \\
    60 & 2514.137 &   51.699 &  114.550421751 &  38.859339644 &   24.490 &  0.039 &  0.057 &   24.442 &  0.038 &  0.093 &   24.205 &  0.053 &   24.333 &  0.039 &   24.468 &  0.039 &   24.949 &  0.027 &   25.305 &  0.083 &   1 1   \\
    63 & 3018.596 &   51.412 &  114.544567990 &  38.856049555 &   25.883 &  0.098 &  0.271 &   25.988 &  0.102 &  0.538 &   25.621 &  0.117 &   25.644 &  0.072 &   25.639 &  0.121 &   26.114 &  0.072 &   26.430 &  0.128 &   1 1   \\
    65 & 3191.724 &   52.050 &  114.542548373 &  38.854924388 &   25.240 &  0.064 &  0.102 &   25.142 &  0.059 &  0.322 &   25.063 &  0.069 &   25.248 &  0.056 &   25.320 &  0.060 &   25.890 &  0.081 &   26.041 &  0.123 &   1 1   \\
    82 & 2679.410 &   52.693 &  114.548495944 &  38.858272263 &   24.676 &  0.044 &  0.042 &   24.610 &  0.042 &  0.079 &   24.413 &  0.040 &   24.538 &  0.033 &   24.669 &  0.034 &   25.165 &  0.039 &   25.550 &  0.107 &   1 1   \\
    94 & 2080.912 &   53.448 &  114.555422572 &  38.862170333 &   25.139 &  0.065 &  0.144 &   25.012 &  0.056 &  0.354 &   25.080 &  0.072 &   25.171 &  0.041 &   25.157 &  0.040 &   25.673 &  0.044 &   26.128 &  0.117 &   1 1   \\
    98 & 3132.494 &   53.630 &  114.543223253 &  38.855325091 &   24.977 &  0.051 &  0.068 &   24.942 &  0.056 &  0.164 &   24.668 &  0.051 &   24.901 &  0.039 &   25.014 &  0.041 &   25.456 &  0.033 &   25.667 &  0.090 &   1 1   \\
    \hline
\end{tabular}
}
\tablefoot{X and Y are the coordinates on the WFC3 detector. RA and DEC are the J2000.0 equatorial coordinates in degrees.
The columns F336W, F343N, F438W, etc., indicate the apparent magnitude in the corresponding filter (STMAG), and eF336W, eF343N, eF438W, etc., the associated photometric errors. For F336W and F343N, rF343N and rF336W give the r.m.s. of the magnitudes measured on the individual frames. The column labelled ``Det'' indicates the detectors on which the star was imaged in the GO-15078 and GO-11903 datasets. The full table is available on-line.}
\end{table*}

\end{appendix}

\end{document}